\documentclass[12pt]{article}
\pdfoutput=1
\usepackage[bookmarks=true,hyperfigures=true]{hyperref}
\usepackage{amsfonts}
\usepackage[dvips]{graphicx}
\usepackage{amsmath,amssymb,mathtools,slashed}
\usepackage[nosort]{cite}
\usepackage[bulletsep]{collref}
\usepackage{color}

\usepackage[margin=2.5cm]{geometry}

\newcommand{\be}{\begin{eqnarray}}
\newcommand{\ee}{\end{eqnarray}}
\newcommand{\la}{\lambda}
\newcommand{\NN}{\mathcal{N}}
\newcommand{\CC}{\mathcal{C}}
\newcommand{\GG}{\mathcal{G}}
\newcommand{\OO}{\mathcal{O}}
\newcommand{\s}{\sigma}
\newcommand{\half}{\sfrac{1}{2}}
\newcommand{\al}{\alpha}
\newcommand{\p}{\partial}
\newcommand{\tQ}{\widetilde Q}
\newcommand{\tS}{\widetilde S}
\newcommand{\dda}{{\dot{\alpha}}}

\newcommand{\ve}{\varepsilon}
\newcommand{\tM}{{\tilde M}}
\newcommand{\tN}{{\tilde N}}
\newcommand{\tL}{{\tilde L}}
\newcommand{\slk}{\slashed k}
\newcommand{\slpsi}{\slashed\psi}
\newcommand{\eps}{\varepsilon}

\newcommand{\nn}{\nonumber}

\newcommand{\G}{\Gamma}
\newcommand{\hE}{\hat\eta}
\newcommand{\hG}{\hat\Gamma}
\newcommand{\op}[1]{\mathcal{#1}}
\newcommand{\transpose}{^\mathsf{T}}
\newcommand{\alphap}{\alpha^\prime}

\DeclareMathOperator{\diag}{diag}

\numberwithin{equation}{section}

\providecommand{\hypersetup}[1]{}
\providecommand{\hyperref}[2][]{#2}
\providecommand{\texorpdfstring}[2]{#1}
\providecommand{\pdfbookmark}[3][]{}
\hypersetup{plainpages=false}
\hypersetup{pdfpagemode=UseOutlines}
\hypersetup{bookmarksnumbered=true}
\hypersetup{bookmarksopen=true}
\hypersetup{pdfstartview=FitH}
\hypersetup{colorlinks=false}
\hypersetup{citebordercolor={.6 .9 .6}}
\hypersetup{urlbordercolor={.7 .8 1}}
\hypersetup{linkbordercolor={1 .7 .7}}
\hypersetup{pdfborder={0 0 .5}}

\makeatletter
\let\@keywords\@empty
\let\@subject\@empty
\providecommand{\keywords}[1]{\gdef\@keywords{#1}}
\providecommand{\subject}[1]{\gdef\@subject{#1}}
\def\thetitle{\@title}
\def\theauthor{\@author}
\def\thesubject{\@subject}
\def\thedate{\@date}
\def\thekeywords{\@keywords}
\makeatother
\AtBeginDocument{
\hypersetup{pdftitle={\thetitle}}%
\hypersetup{pdfauthor={\theauthor}}%
\hypersetup{pdfsubject={\thesubject}}%
\hypersetup{pdfkeywords={\thekeywords}}%
}

\makeatletter
\def\mr@ignsp#1 {\ifx\:#1\@empty\else #1\expandafter\mr@ignsp\fi}%
\newcommand{\multiref}[1]{\begingroup
\xdef\mr@no@sparg{\expandafter\mr@ignsp#1 \: }%
\def\mr@comma{}%
\@for\mr@refs:=\mr@no@sparg\do{\mr@comma\def\mr@comma{,}\ref{\mr@refs}}%
\endgroup}
\makeatother

\renewcommand{\eqref}[1]{(\multiref{#1})}

\newcommand{\namedref}[2]{\hyperref[#2]{#1~\ref{#2}}}
\newcommand{\secref}[1]{\namedref{Section}{#1}}
\newcommand{\appref}[1]{\namedref{Appendix}{#1}}

\let\oldbfseries=\bfseries
\let\oldmdseries=\mdseries
\let\oldnormalfont=\normalfont
\renewcommand{\bfseries}{\oldbfseries\boldmath}
\renewcommand{\mdseries}{\oldmdseries\unboldmath}
\renewcommand{\normalfont}{\oldnormalfont\unboldmath}

\newcommand{\bfm}[1]{\boldsymbol{#1}}

\newcommand{\earel}[1]{\mathrel{}&\hspace{-2\arraycolsep}#1\hspace{-2\arraycolsep}&\mathrel{}}
\newcommand{\eq}{\earel{=}}
\newcommand{\eqv}{\earel{\equiv}}
\newcommand{\appr}{\earel{\approx}}

\allowdisplaybreaks[4]

\DeclareMathSymbol{\Gamma}{\mathalpha}{letters}{"00}
\DeclareMathSymbol{\Delta}{\mathalpha}{letters}{"01}
\DeclareMathSymbol{\Theta}{\mathalpha}{letters}{"02}
\DeclareMathSymbol{\Lambda}{\mathalpha}{letters}{"03}
\DeclareMathSymbol{\Xi}{\mathalpha}{letters}{"04}
\DeclareMathSymbol{\Pi}{\mathalpha}{letters}{"05}
\DeclareMathSymbol{\Sigma}{\mathalpha}{letters}{"06}
\DeclareMathSymbol{\Upsilon}{\mathalpha}{letters}{"07}
\DeclareMathSymbol{\Phi}{\mathalpha}{letters}{"08}
\DeclareMathSymbol{\Psi}{\mathalpha}{letters}{"09}
\DeclareMathSymbol{\Omega}{\mathalpha}{letters}{"0A}

\newcommand{\brk}[1]{(#1)}
\newcommand{\lrbrk}[1]{\left(#1\right)}
\newcommand{\bigbrk}[1]{\bigl(#1\bigr)}
\newcommand{\biggbrk}[1]{\biggl(#1\biggr)}
\newcommand{\Bigbrk}[1]{\Bigl(#1\Bigr)}

\newcommand{\bigbrc}[1]{\bigl\{#1\bigr\}}

\newcommand{\bigsbrk}[1]{\bigl[#1\bigr]}
\newcommand{\biggsbrk}[1]{\biggl[#1\biggr]}
\newcommand{\Bigsbrk}[1]{\Bigl[#1\Bigr]}
\newcommand{\abs}[1]{|#1|}
\newcommand{\vev}[1]{\langle#1\rangle}
\newcommand{\bigvev}[1]{\bigl\langle#1\bigr\rangle}
\newcommand{\Bigvev}[1]{\Bigl\langle#1\Bigr\rangle}

\newcommand{\normord}[1]{\mathopen{:}#1\mathclose{:}}
\newcommand{\comm}[2]{[#1,#2]}
\newcommand{\bigcomm}[2]{\big[#1,#2\big]}

\newcommand{\Bigcomm}[2]{\Big[#1,#2\Big]}

\newcommand{\bigacomm}[2]{\big\{#1,#2\big\}}

\newcommand{\grp}[1]{\mathrm{#1}}
\newcommand{\alg}[1]{\mathfrak{#1}}
\newcommand{\indup}[1]{_{\mathrm{#1}}}

\newcommand{\supup}[1]{^{\mathrm{#1}}}

\newcommand{\sfrac}[2]{{\textstyle\frac{#1}{#2}}}

\newcommand{\gc}{g\indup{c}}
\newcommand{\tw}{\Theta} 
\newcommand{\tr}{\mathop{\mathrm{Tr}}}
\newcommand{\Tr}{\mathop{\mathrm{Tr}}}
\newcommand{\order}[1]{\mathcal{O}(#1)}
\newcommand{\superN}{\mathcal{N}}
\newcommand{\sym}{$\superN=\nolinebreak4$ SYM}
\newcommand{\scdot}{\!\cdot\!}
\newcommand{\ap}{\alpha'}
\newcommand{\apt}{\frac{\alpha'}{2}}
\newcommand{\sapt}{\sfrac{\alpha'}{2}}

\newcommand{\stap}{\sfrac{2}{\alpha'}}

\title{Computing Three-Point Functions for Short Operators}
\author{Till Bargheer, Joseph A. Minahan, Raul Pereira}
\subject{Gauge Theory, String Theory, AdS/CFT}
\keywords{vertex operators, worldsheet theory, correlation functions, three-point functions, strong coupling, primaries, level one}

\begin{document}

\pdfbookmark[1]{Title Page}{title}

\thispagestyle{empty}
\setcounter{page}{0}
\begin{flushright}\footnotesize
\texttt{DESY 13-201}
\\\texttt{UUITP-17/13}
\end{flushright}

\vspace{0.5cm}

\begin{center}
{\Large\textbf{Computing Three-Point Functions\\for Short Operators}\par}

\vspace{12mm}

\small

\textsc{Till Bargheer,${}^{1,2}$ Joseph~A.~Minahan${}^3$ and Raul Pereira${}^3$}

\bigskip

\textit{${}^1$%
School of Natural Sciences, The Institute for Advanced Study\\
Einstein Drive, Princeton, NJ 08540, USA}

\medskip

\textit{${}^2$%
DESY Theory Group, DESY Hamburg\\
Notkestra{\ss}e 85, D-22603 Hamburg, Germany}

\medskip

\textit{${}^3%
$Department of Physics and Astronomy, Uppsala University\\
Box 520, SE-751 20 Uppsala, Sweden}

\bigskip

\texttt{bargheer@ias.edu, joseph.minahan \& raul.pereira@physics.uu.se}

\normalsize

\vspace{15mm}

\textbf{Abstract}

\vspace{5mm}

\begin{minipage}{12cm}

We compute the three-point
structure constants
for short primary operators of
$\superN=4$ super Yang--Mills theory to
leading order in $1/\sqrt{\lambda}$ by mapping the problem to a
flat-space string theory calculation.  We check the validity of our
procedure by comparing
to known results for three chiral primaries.
We then compute the three-point
functions for any combination of chiral and non-chiral primaries,
with the non-chiral primaries all dual to string states at the first
massive level.  Along the way we find many cancellations that leave us
with simple expressions, suggesting that integrability is playing an
important role.

\end{minipage}

\end{center}

\newpage

\renewcommand{\thefootnote}{\arabic{footnote}}

\setcounter{tocdepth}{3}
\hrule height 0.75pt
\pdfbookmark[1]{\contentsname}{contents}
{\small\tableofcontents}
\vspace{0.8cm}
\hrule height 0.75pt
\vspace{1cm}

\section{Introduction}

It is now possible, at least in principle, to compute the dimensions
of single-trace local operators in planar $\NN=4$ super Yang--Mills for any value of the 't~Hooft coupling $\lambda$.  The ability to do this relies on the underlying integrability of the theory~\cite{Beisert:2010jr}.  One of the most impressive results is the numerical computation of the dimension, $\Delta\indup{K}$, of the Konishi operator, $\OO\indup{K}=\tr\Phi^I\Phi_I$.  Starting at weak coupling and interpolating
to very large values of $\lambda$,
one finds extremely precise results~\cite{Gromov:2009zb,Frolov:2010wt}.
What is clear from these studies is that for large values of
$\lambda$, $\Delta\indup{K}$ asymptotes to
\be\label{Konishidim}
\Delta\indup{K}=2\lambda^{1/4}-2+\frac{2}{\lambda^{1/4}}+\dots\qquad\lambda\gg1\,,
\ee
a very satisfying result since the leading behavior was predicted well beforehand based on a flat-space approximation of a string on $\grp{AdS}_5\times\grp{S}^5$~\cite{Gubser:1998bc}.  The next two terms in this series can also be motivated by semi-classical string theory computations~\cite{Roiban:2009aa,Gromov:2011de,Beccaria:2011uz,Roiban:2011fe,Beccaria:2012xm,Beccaria:2012kp}, or by analyzing the algebraic curve~\cite{Basso:2010in,Gromov:2011bz}, or by a mini-superspace approach on the string world-sheet~\cite{Frolov:2013lva}.

To fully solve planar $\NN=4$ SYM it is also necessary to know the three-point functions between local operators.  Compared to the spectrum, the results here are rather limited.  However, there are indications that integrability will play a crucial role in determining the structure constants. In particular, there have been important results at low orders of perturbation theory that crucially rely on the integrable structure in
the
theory~\cite{Escobedo:2010xs,Escobedo:2011xw,Gromov:2011jh,Kostov:2012yq,Foda:2013nua}.
Impressively, in~\cite{Escobedo:2011xw} it was shown that the
weak-coupling three-point function for two semi-classical operators
(with $\Delta\sim\sqrt{\lambda}$)
and one short BPS operator, was consistent with the strong-coupling results derived from a semi-classical string calculation in~\cite{Zarembo:2010rr,Costa:2010rz}.

Based on our experience with two-point functions, it will be much more
difficult to compare weak and strong coupling results if the
operators in the three-point functions are all short.%
\footnote{By
``short" we mean local single-trace operators whose dual string
states are short compared to the radii of curvature in
$\grp{AdS}_5\times\grp{S}^5$. We do not mean operators that are in
short super-multiplets.}
This is because finite-size effects are
expected to become important at strong-coupling and so all of the
machinery of the thermodynamic Bethe ansatz
will likely be needed.  Nonetheless, as with $\Delta\indup{K}$, it will be very useful to know the strong-coupling target.  In other words, analogous to the GKP result in~\cite{Gubser:1998bc}, we wish to know the leading strong-coupling behavior for such three-point functions.

In this paper we continue the study of three-point functions for
short operators
at level one.
The methods we describe take advantage of the fact that at large
Yang--Mills coupling the string theory target space has small
curvature and so one can approximate the vertex operators with
flat-space vertex operators~\cite{Minahan:2012fh}.  The curvature
still plays a role as it determines how the states propagate in from
the boundary.  The accompanying Witten diagrams have an intersection
point that should be integrated over, but for short operators with a
large dimension the integral is dominant over a small region of the
bulk%
\footnote{The same thing leads to the focusing of wave-packets for AdS $S$-matrices~\cite{Polchinski:1999ry,Susskind:1998vk,Balasubramanian:1999ri}.}
where one can ignore the curvature as well as the Ramond--Ramond flux.  Within that region, one can then use flat space vertex operators to compute the couplings that determine the part of the three-point function not determined solely by the dimensions of the operators.

The challenge is to find the flat-space vertex operators and then
compute the three-point function with those respective operators.
In~\cite{Minahan:2012fh} it was described how to find such operators.
In particular, it was argued that the relevant operators satisfy a
twisted version of \mbox{$Q_{\mathrm{L}}=i\,Q_{\mathrm{R}}$}, where
$Q_{\mathrm{L,R}}$ are the left and right ten-dimensional supercharges in flat space.  It was shown that this condition on the supercharges maps back to the AdS condition that $S=0$, where $S$ represents all superconformal charges.  We impose this AdS condition since the three operators are assumed to be primaries.  Consequently, the operators are necessarily combinations of NS-NS and R-R scalars.

Since the Konishi operator is primary but not chiral primary, its string dual is a massive state.  In the large coupling limit, the dimensions of scalar operators with no $R$-charges have dimensions that are approximately $\Delta\approx2\sqrt{n}\,\la^{1/4}$ corresponding to the energies of stringy modes when $\ap=1/\sqrt{\la}$~\cite{Gubser:1998bc}. There is now overwhelming evidence that the Konishi operator approaches one of these stringy modes with $n=1$~\cite{Bianchi:2003wx,Beisert:2003te,Gromov:2009zb}\nocollect{Gromov:2009zb}.  Hence, the three-point function for three Konishi operators at strong coupling can be well-approximated by determining the three-point amplitude for three massive string states in flat IIB string theory.

Unfortunately, the vertex operators are quite complicated, even at the
lowest massive level, and to the best of our knowledge a three-point
function involving three massive IIB vertex operators has never been
computed.%
\footnote{The form of superstring three-point amplitudes are
constrained by supersymmetric Ward and factorization identities
\cite{Boels:2012if}. It would be interesting to understand to what
extent such identities can determine the R-charge-dependent normalization
factors that we compute in this work.}
In this paper we will do just that for three scalar
operators, although they are not scalars in the ten-dimensional
theory.  We will also compute three-point functions involving one, two
and three chiral primaries, which are dual to massless string states,
and whose dimensions equal the sum of their $R$-charges.  In the case
of three chiral primaries we will compare our result to the
supergravity result in~\cite{Lee:1998bxa}, showing that our procedure
is consistent with their result.  The correlator with one chiral
primary whose $R$-charge $J$ satisfies $J\ll\lambda^{1/4}$ is of
significance because it comes with an exponential suppression factor that depends on $J$ and not on the dimensions of the two non-chiral primaries.

The full three-point function of the operators also relies on overlap
integrals on the $\grp{S}^5$.  For chiral primaries, it is necessary
to assume that the $R$-charges are large so that their corresponding
dimensions are large.  This will result in wave-functions that are
strongly peaked on $\grp{S}^5$.  However, for non-chiral primaries the
dimensions are already large in the strong-coupling limit and there is
otherwise no reason to assume large $R$-charges for these
operators,%
\footnote{We thank E. Witten for a discussion on this
point.}
provided that the total $R$-charge in the three-point function
is conserved.   In any case we will give expressions for three-point
functions for general $J$ values,
but also give results for when the $R$-charges of the
non-chiral primaries are
zero, which happens for three Konishi operators.

The three-point function for three scalar operators in $d$ space-time
dimensions with scaling dimensions $1\ll\Delta_i\ll\sqrt{\lambda}$ is given by
\be\label{3corr1}
\bigvev{\OO_{\Delta_1}(x^\mu_1)\OO_{\Delta_2}(x^\mu_2)\OO_{\Delta_3}(x^\mu_3)}\eq\frac{
\mathcal{C}_{123} }{|x_{12}|^{\Delta_1+\Delta_2-\Delta_3}|x_{23}|^{\Delta_2+\Delta_3-\Delta_1}|x_{31}|^{\Delta_3+\Delta_1-\Delta_2}}\,,
\ee
where~\cite{Minahan:2012fh,Klose:2011rm}
\be\label{3corr2}
\mathcal{C}_{123}\approx\frac{\pi^{\frac{2-d}{4}}}{4}\,\frac{(\Delta_1\Delta_2\Delta_3)^{d/4}}{\left(\al_1\al_2\al_3\Sigma^{d+1}\right)^{1/2}}\,\frac{\al_1^{\al_1}\al_2^{\al_2}\al_3^{\al_3}\Sigma^\Sigma}{\Delta_1^{\Delta_1}\Delta_2^{\Delta_2}\Delta_3^{\Delta_3}}\,\GG_{123}
\ee
and
\begin{gather}
\al_1=\half(\Delta_2+\Delta_3-\Delta_1)\,,
\qquad
\al_2=\half(\Delta_3+\Delta_1-\Delta_2)\,,
\qquad
\al_3=\half(\Delta_1+\Delta_2-\Delta_3)\,,
\nn\\
\Sigma=\half(\Delta_1+\Delta_2+\Delta_3)=\al_1+\al_2+\al_3\,.
\label{eq:alphasigma}
\end{gather}
The coupling $\GG_{123}$ is given by
\be\label{coupling}
\GG_{123}=
\frac{8\pi}{\gc^2\ap}\,\vev{V_{k_1}V_{k_2}V_{k_3}}\,\vev{\psi_{J_1}\psi_{J_2}\psi_{J_3}}\,,
\ee
which includes the contribution of the overlap integral
$\vev{\psi_{J_1}\psi_{J_2}\psi_{J_3}}$ on $S^5$.
The closed string coupling $\gc$ and the inverse string tension are
given under the AdS/CFT dictionary as \mbox{$\gc=\pi^{3/2}/N$} and $\ap=1/\sqrt{\lambda}$.

In this paper we will find for three level-one states with
small $R$-charges $J_i\ll\lambda^{1/4}$
that the flat-space vertex operators satisfy
\be\label{result}
\vev{V_{k_1}V_{k_2}V_{k_3}}
=
\gc^3\,\frac{3^8}{2^9}
+\order{J_i^2\lambda^{-1/2}}.
\ee
Hence the result for the correlator of three
short operators with small $R$-charges reads
\be\label{CCresult}
\CC_{123}
\approx
\frac{1}{N}
\,\frac{ 3^{11/2}\,{\pi^2}}{32}
\lambda^{1/4}
\lrbrk{\frac{3}{4}}^{3\Delta/2}
\vev{\psi_{J_1}\psi_{J_2}\psi_{J_3}}\,.
\ee
For vanishing $R$-charges $J_i$ we have that
$\vev{\psi_{J_1}\psi_{J_2}\psi_{J_3}}=\pi^{-3/2}$.
Including the finite correction term in \eqref{Konishidim}~\cite{Roiban:2009aa},
we can thus re\"express \eqref{CCresult} for three Konishi operators as
\be\label{CCresulta}
\CC_{123}
\approx
\frac{1}{N}
\,64\,{\pi^{1/2}}
\lambda^{1/4}
\lrbrk{\frac{3}{4}}^{3\lambda^{1/4}+5/2}
\,.
\ee
In the intermediate steps  the individual contributions to \eqref{result} involve larger prime factors, but after combining the different parts we end up with a much simpler expression.
This suggests that there is an underlying symmetry playing an important role.

For the case of one chiral primary and two primaries at level one, we find that
\be\label{CCresult2}
\CC_{123}\approx\,\frac{1}{N}\,\pi^2\,\lambda^{1/4}\,2^{-J_3}\;\vev{\psi_{J_1}\psi_{J_2}\psi_{J_3}}\,,
\ee
where $1\ll J_3\ll\lambda^{1/4}$ is the magnitude of the $R$-charge of
the chiral primary, and $\vev{\psi_{J_1}\psi_{J_2}\psi_{J_3}}$ depends
on the distribution of $R$-charge on the level-one states. For
$J_1=0$, $J_2=J_3=J$,
the correlator becomes
\begin{equation}
\CC_{123}\approx\,\frac{1}{N}\,\pi^{1/2}\,\lambda^{1/4}\,2^{-J}\,.
\end{equation}
Note that
$\CC_{123}$ is exponentially suppressed if all $\Delta_i\gg1$.  However, if one of the operators is a chiral primary with $\Delta=J$ and $J$ is much smaller than the other two dimensions, then the suppression factor is only by $2^{-J}$.

The rest of this paper is organized as follows: In \secref{sec:review} we review some results of~\cite{Minahan:2012fh}.  In
\secref{sec:results}, which is the main part of this article, we compute
three-point amplitudes for three string states where zero, one, two
or three are at the first massive level.  We then use these to reach
the main results in \eqref{CCresult} and \eqref{CCresult2}.  The
partial amplitudes involve combinations of three NS-NS scalars or one NS-NS scalar and two R-R
scalars.  The states are not Lorentz scalars in the full
ten-dimensional flat space, but instead are twisted versions of these
states.  However, for $J=0$ they remain scalars in the two reduced
five-dimensional spaces which are the flattened versions of AdS$_5$
and $\grp{S}^5$.%
\footnote{They are no longer AdS scalars if $J\ne 0$ because the
twisting happens in the four-dimensional subspace transverse to the
AdS momentum, but the polarization can have components along both the
AdS momentum and its transverse space.}
\secref{sec:conclusions} contains our conclusions and discussion for
various extensions of these results.  Many technical details are
contained in several appendices.

\section{Review of Previous Results}
\label{sec:review}

In this section we collect some relevant results from~\cite{Minahan:2012fh}.

We assume that the operators are short operators in the sense that
their sizes are much smaller than the AdS$_5$ and S$^5$ radii. Hence at scales comparable to this radius the strings appear point-like.  The three-point correlator is then closely related to the computation of a three-point correlator using Witten diagrams~\cite{Witten:1998qj,Freedman:1998tz}.

In the Witten diagram one integrates over all possible intersection
points, but in the semiclassical approximation the Witten diagrams are
dominated by peaks along the particles' classical trajectories in the
$\grp{AdS}_5\times\grp{S}^5$ background.  The intersection point is
calculable~\cite{Klose:2011rm,Minahan:2012fh}, and after including the
contributions of the fluctuations one finds the coefficient in front
of $\GG_{123}$ in \eqref{3corr2}.  For large $R$-charges $J_i\gg 1$ we can also use a semiclassical approximation for the $\grp{S}^5$ overlap integrals $\vev{\psi_{J_1}\psi_{J_2}\psi_{J_3}}$~\cite{Buchbinder:2011jr,Minahan:2012fh}.  In this approximation we assume that the $R$-charges are each highest weight with respect to some basis and satisfy $\vec J_1+\vec J_2+\vec J_3=0$.  Then one finds
\be
\vev{\psi_{J_1}\psi_{J_2}\psi_{J_3}}\appr\frac{1}{2\pi^2}\frac{(J_1J_2J_3)^{3/2}}{\brk{\tilde\al_1\tilde\al_2\tilde\al_3\tilde\Sigma^5}^{1/2}}\,,\ee
where
\be\label{tildedefs}
\tilde\Sigma=\frac{J_1+J_2+J_3}{2}\,,
\quad
\tilde\al_1=\frac{J_2+J_3-J_1}{2}\,,
\quad
\tilde\al_2=\frac{J_3+J_1-J_2}{2}\,,
\quad
\tilde\al_3=\frac{J_1+J_2-J_3}{2}\,.
\ee
For later reference, these can be written in terms of the standard spherical harmonic coefficients $C^J_{I_1\dots I_J}$ in~\cite{Lee:1998bxa} as
\be\label{3S5}
\vev{\psi_{J_1}\psi_{J_2}\psi_{J_3}}\appr\frac{1}{2\pi^2}\frac{(J_1J_2J_3)^{3/2}}{(\tilde\al_1\tilde\al_2\tilde\al_3\tilde\Sigma^5)^{1/2}}\frac{J_1^{J_1}J_2^{J_2}J_3^{\tilde J_3}}{\tilde\al_1^{\tilde\al_1}\tilde\al_2^{\tilde\al_2}\tilde\al_3^{\tilde\al_3}\tilde\Sigma^{\tilde\Sigma}}\vev{C^{J_1}C^{J_2}C^{J_3}}\,,
\ee
where
\be\label{CCC}
\vev{C^{J_1}C^{J_2}C^{J_3}}
\approx
\frac{
\tilde\al_1^{\tilde\al_1}\tilde\al_2^{\tilde\al_2}\tilde\al_3^{\tilde\al_3}\tilde\Sigma^{\tilde\Sigma}
}{J_1^{J_1}J_2^{J_2}J_3^{J_3}}\,.
\ee

The classical trajectories come with constants of motion which are
determined by the $R$-charges, the spins, the dimensions and by the
positions of the operators on the boundary.  The constants of the motion can be formulated in terms of the components of the conformal algebra, $P_\mu$, $K^\mu$, $D$ and $M_{\mu\nu}$.  For the operator $\OO_i(x_i^\mu)$ in the three-point function the conserved charges include
\be
\vev{P^\mu_{i}}\eq-2i\sum_{\mathclap{i\neq j\neq k\neq i}}\alpha_k\frac{x_{ij}^\mu}{x_{ij}^2}\,,\nn\\
\vev{D_i}
\eq i\biggbrk{\Delta_i-2\,x_{i\mu}\sum_{\mathclap{i\neq j\neq k\neq i}}\alpha_k\frac{x_{ij}^\mu}{x_{ij}^2}}
\,,
\ee
where $x^\mu_{ij}=x^\mu_i-x^\mu_j$.  The sums are over the other two
operators in the three-point function.  One can also construct the
quadratic Casimir operator
\be
-2\vev{P_\mu}\vev{K^\mu}+\vev{D}^2-\frac{1}{2}\vev{M_{\mu\nu}}\vev{M^{\mu\nu}}=-\Delta^2\,.
\ee

At the point where the three trajectories intersect these constants of the motion are conserved.
It is convenient to shift the operator positions such that the intersection point is at $x^\mu=0$.  In this case all $\vev{M_{\mu\nu}}=0$ and the conserved charges $\vev{K^\mu_i}$ satisfy
\be\label{KPrel}
\vev{K_i^\mu}=-\frac{\al_1\al_2\al_3\Sigma}{F^2}x_{12}^2x_{23}^2x_{31}^2\vev{P_i^\mu}\,,
\ee
where
\be
F=\al_1\al_2\, x_{12}^2+\al_2\al_3\,x_{23}^2+\al_3\al_1\,x_{13}^2\,.
\ee
It is also convenient to write everything in a manifest
$\grp{SO}(2,d)$
formalism, with
\be\label{so2d}
M_{-1\mu}\equiv\sfrac{1}{\sqrt{2}}(\kappa P_\mu-\kappa^{-1}K_\mu)\qquad M_{d\mu}\equiv\sfrac{1}{\sqrt{2}}(\kappa P_\mu+\kappa^{-1}K_\mu)\qquad M_{-1d}\equiv -D\,,
\ee
where $\kappa$ is arbitrary.  In this convention the Casimir operator
becomes $-\half \vev{M_{rs}}\vev{M^{rs}}=-\Delta^2$, $r,s=-1\dots d$. Setting
$d=4$, we can also
combine this with the $\grp{SO}(6)$ $R$-symmetry Casimir to write
\be
-\half \vev{M_{rs}}\vev{M^{rs}}+\half \vev{R_{IJ}}\vev{R^{IJ}}=-\Delta^2+J^2\,,
\ee
where $R_{IJ}$ are the $R$-symmetry generators with
$I,J=5,\dots,10$.

If $\Delta$ and $J$ are large, then the $\grp{SO}(2,4)\times
\grp{SO}(6)$ algebra effectively reduces to a 10 dimensional Poincar\'e algebra.
Assuming that all $\vev{M_{\mu\nu}}=0$,
and choosing $\kappa$ to be
\be
\kappa =\frac{\sqrt{\al_1\al_2\al_3\Sigma}}{F}|x_{12}||x_{23}||x_{31}|\,,
\ee
then the only non-zero components in \eqref{so2d} are $\vev{M_{-1m}}$,
$m=0,\dots,4$
for all three operators.  Assuming $\Delta^2\gg1$, and choosing a
basis where the only non-zero $R$-symmetry components are $\vev{R_{J,10}}$, we can then identify the full 10-dimensional flat-space momentum as
\be\label{10dmom}
k^M=\bigbrk{\vev{M_{-1m}},\vev{R_{J,10}}}\,,
\ee
which has to satisfy the on-shell condition
\be\label{k2}
k\cdot k=-\Delta^2+J^2=-4n\sqrt{\la}\,.
\ee

If we fill out the algebra to the full superconformal $\grp{PSU}(2,2|4)$,
then we also have fermionic generators.  These include the
supergenerators $Q_{\al a}$ and $\tQ^a_{\dda}$, where $\al$ and
$\dda$ are space-time
spinor indices and $a$ is an $\grp{SO}(6)$ spinor index
(raised or lowered depending on which spinor representation), and the
superconformal generators $S^a_{\al}$ and $\tS_{\dda a}$.  We can put
these into a form that is
manifestly $\grp{SU}(2,2)\simeq\grp{SO}(2,4)$ covariant by
defining
\be
Q^1_{\dot a a}\eqv(\kappa^{1/2}Q_{\al a},\kappa^{-1/2}\tilde S_{\dot\al a})\nn\\
Q^{2,\dot a a}\eqv(\kappa^{-1/2}\eps^{\al\beta}S^a_{\beta},\kappa^{1/2}\eps^{\dot\al\dot\beta}\tilde Q^a_{\dot\beta})
\ee
where the lowered $\dot a$ is an $\grp{SO}(2,4)$ spinor index and the raised $\dot a$ is an index for the other spinor representation.  The $\kappa$ is the same that appears in \eqref{so2d}.  If we then define two sets of supercharges
\be\label{QLQRdef}
Q^{\mathrm{L}}_A=Q^1_{\dot a a}+\gamma^{-1}_{\dot b\dot a}\gamma^{\ 6}_{ba}Q^{2,\dot bb}\,,\qquad
Q^{\mathrm{R}}_A=-i\bigbrk{Q^1_{\dot a a}-\gamma^{-1}_{\dot b\dot a}\gamma^{\ 6}_{ba}Q^{2,\dot bb}}\,,
\ee
then in the flat-space limit these approach the usual 10-dimensional
super-Poincar\'e generators with
\begin{equation}
\bigacomm{Q\supup{L,R}_A}{Q\supup{L,R}_B}
=
-2(\mathrm{P_+}\Gamma^M C)_{AB}P_M\,,
\qquad
P_M=(M_{-1m},R_{J,10})\,,
\qquad
\bigacomm{Q\supup{L}_A}{Q\supup{R}_B}
=
0
\label{eq:poincare}
\end{equation}
for $\Delta\gg1$, $J\gg1$, where $\mathrm{P_+}$ is the
positive-chirality projector.%
\footnote{For $J\sim\order{1}$,  the correction terms in \eqref{eq:poincare} are of the same order as the $P_J$ components.  However, as long as $\Delta\gg1$ these are suppressed compared to the other momentum components 
and hence do not affect the conditions obeyed by the vertex operators.}

For a primary operator we have that
$[S^a_\al,\OO(0)]=[\tS_{\dda a},\OO(0)]=0$.
If we compare this to the new
generators in \eqref{QLQRdef} we see
that this corresponds to
\be
Q\supup{L}_{A}=\pm i\,Q\supup{R}_{A}
\ee
for the primary operators, where the sign depends on the spinor component. With the operator at $x_i^\mu=0$ it is clear that the operator's charges have $\vev{K^\mu}=\vev{M_{\mu\nu}}=0$.  If we then assume that the intersection point is arbitrarily far away, or equivalently we are on a part of the trajectory arbitrarily close to the boundary, then for finite $\kappa$ we have $M_{-1\mu}=M_{4\mu}=0$, $M_{-1 4}=-i\Delta$, which puts the space-time directions transverse to the momentum in AdS$_5$.  Going back to \eqref{QLQRdef}, we see that in the flat-space limit this imposes the condition on the operators
\be\label{twistcond}
Q\supup{L}_{\al \tilde a}=i\,Q\supup{R}_{\al \tilde a}\,,
\qquad
{Q\supup{L}_{\dot\al}}^{\tilde a}=-i\,{Q\supup{R}_{\dot\al}}^{\tilde a}\,,
\ee
where we use here the explicit four-dimensional space-time
($\al,\dot\al$) and orthogonal six-dimen\-sional ($\tilde a$) spinor
indices. Shifting the intersection point back to $x^\mu=0$, the
$M_{-1\mu}$ are no longer zero, but the primary operator condition is
still defined by the directions transverse to the momentum in AdS$_5$.
Hence, the four-dimensional spinor indices in \eqref{twistcond} are
replaced with the spinor indices for the four-dimensional space
transverse to the particle momentum in AdS$_5$, while the
six-dimensional spinor indices are replaced accordingly.

We then proceed by finding operators that satisfy the condition in
\eqref{twistcond}.  As was emphasized in~\cite{Minahan:2012fh}, this
leads to string states that are linear combinations of NS-NS and R-R
modes.  At the massless level this gives chiral primary operators
while at the massive levels the states correspond only to primary
operators.  The relative sign in \eqref{twistcond} complicates matters
as the states cannot be scalars in the full 10-dimensional flat-space.
At best they can be scalars in the two five-dimensional subspaces.
Nonetheless, it is still convenient to start with twisted versions of
our vertex operators that correspond to 10-dimensional scalars, and
then untwist them afterwards.  In particular, if we start with a vertex
operator $V_{\mathrm{T},k}$ that satisfies
\be\label{twvc}
[Q\supup{L}_{A},V_{\mathrm{T},k}]=[i\,Q\supup{R}_{A},V_{\mathrm{T},k}]\,,
\ee
then this is related to an operator $V_k$ that satisfies \eqref{twistcond} by
\be
V_{\mathrm{T},k}=e^{\pi i(M\supup{R}_{0'1'}+M\supup{R}_{2'3'})}V_k\,,
\ee
where the $\mu'$ components are transverse to $k^M$ in the AdS$_5$
part and $M\supup{R}_{\mu'\nu'}$ is a generator of rotations for the right-movers.

For the level-zero vertex operators, those corresponding to the massless states, a general NS-NS vertex operator in the $(-1,-1)$ picture has the form~\cite{Friedan:1985ge}
\be
W_{1,k}^{(-1,-1)}(z,\bar z)=\gc\,\ve_{M\tilde M}\,\psi^Me^{-\phi}\tilde\psi^{\tilde M}e^{-\tilde\phi}e^{ik\cdot X}\,,
\qquad
k^M\ve_{M\tilde M}=k^{\tilde M}\ve_{M\tilde M}=0\,,
\label{eq:W1m1}
\ee
where $\phi$ and $\tilde\phi$ are the left and right superconformal ghost fields.
Eventually we will replace $\gc$ and $\ap$ with their dictionary
values, but for now leave them in the usual string-theory form.   Since the total ghost number for the three-point function must be $(-2,-2)$, we will also need the vertex operator in the $(0,0)$ picture, which is given by
\be
W_{1,k}^{(0,0)}(z,\bar z)
=
-\gc\,\ve_{M\tilde M}\,\stap
\bigbrk{i\p X^M+\sapt k\scdot\psi\psi^M}
\bigbrk{i\bar\p X^{\tilde
M}+\sapt k\scdot\tilde\psi\tilde\psi^{\tilde M}}
e^{ik\cdot X}\,.
\label{eq:W10}
\ee
A generic vertex operator in the R-R sector is~\cite{Friedan:1985ge}
\be
W_{2,k}^{(-1/2,-1/2)}(z,\bar z)=\gc\,t_{AB}\,\tilde\Theta^Ae^{-\half\tilde\phi}\,\Theta^Be^{-\half\phi}e^{ik\cdot X}\,,\quad{t\,\slk=0}\,,
\label{eq:W2}
\ee
where $\Theta_A$ and $\tilde\Theta_B$ are the sixteen-component left and right twist fields.  We only need these vertex operators in the $(-1/2,-1/2)$ picture.

The twisted vertex operator will be a linear combination of the NS-NS and the R-R vertex operators.  Equation \eqref{twvc} can only be satisfied up to spurious terms and has one scalar solution,
\be
W_{\mathrm{T}}=W_{1,\mathrm{T}}+\frac{1}{\sqrt{2}}W_{2,\mathrm{T}}\,,
\label{eq:WT}
\ee
up to a normalization factor.
Here, $W_{1,\mathrm{T}}$ and $W_{2,\mathrm{T}}$ are the NS-NS and R-R
vertex operators with polarizations%
\footnote{The coefficient of the R-R polarization $t_{AB}$ depends on
the chiralities of the spin fields $\tilde\tw^A$, $\tw^B$. Since the worldsheet theory is chiral,
we will  assume throughout the paper that all spin fields
$\tw$, $\tilde\tw$  have positive chirality.}
\be\label{soln=0}
\ve^{M\tilde M}_{\mathrm{T}}=\eta^{M\tilde M}-\frac{k^M\bar k^{\tilde M}+k^{\tilde M}\bar k^M}{k\cdot\bar k}\,,
\qquad\qquad
t_{\mathrm{T},AB}=\left(\sapt\right)^{1/2} (C^\dagger\slk)_{AB}\,,
\ee
where $C$ is the charge conjugation matrix
and
$\bar k^N$ is an arbitrary light-like vector satisfying $k\cdot\bar k\ne 0$.  Untwisting the vertex operators the NS-NS polarization becomes
\be
\ve^{M\tilde M}\eq(-1)^{\s_k(\tilde M)}\ve^{M\tilde M}_{\mathrm{T}}\,
\ee
where
\begin{equation}
\s_k(N)=
\begin{cases}
1 & N=0',\dots,3'\\
0 & N=4',5,\dots,9\,.
\end{cases}
\label{sdef}
\end{equation}
Untwisting the R-R vertex operator, $t_{\mathrm{T},AB}$ gets replaced with
\be
t_{AB}=\left(\sapt\right)^{1/2}
\bigbrk{C^{\dagger}i\Gamma^{0'}\Gamma^{1'}\Gamma^{2'}\Gamma^{3'}\slk}_{AB}\,.
\ee
As before, the primed indices denote a frame in which
the directions $0',\dots,3'$ are transverse to $k$.
The NS-NS vertex operator is for a graviton component in the full 10-dimensional space-time while the R-R vertex operator is that for a self-dual 4-form field.  Further details on the massless vertices can be found in \appref{sec:masslessvert}.

Moving now to level one there are two different types of NS-NS vertex operators that we can construct.  The first has the form
in the $(-1,-1)$ picture~\cite{Koh:1987hm}
\be
V_{1,k}^{(-1,-1)}(z,\bar z)=\gc\stap\eps_{MN;\tM\tN}\,\psi^M i\p X^N e^{-\phi}\,\tilde\psi^{\tM}i\bar\p X^{\tN} e^{-\tilde\phi}e^{ik\cdot X}\,,
\label{eq:V1m1}
\ee
where the first two and last two indices of $\eps_{MN;\tM\tN}$ are symmetric and traceless.
The second vertex in the $(-1,-1)$ picture is given by
\be
V_{2,k}^{(-1,-1)}(z,\bar z)\eq\gc\,\al_{MNL;\tM\tN\tL}\,\psi^M\psi^N\psi^Le^{-\phi}\,\tilde\psi^\tM\tilde\psi^\tN\tilde\psi^\tL e^{-\tilde\phi}e^{ik\cdot X}\,.
\ee
where the first three and last three indices of $\al_{MNL;\tM\tN\tL}$
are antisymmetric.  For physical states the contraction of any index
in $\eps_{MN;\tM\tN}$ or $\al_{MNL;\tM\tN\tL}$ with $k^M$ vanishes.  There are $44\times 44$ independent polarizations for $V_1$ and $84\times 84$ for $V_2$, but
only one of each type is a ten-dimensional Lorentz scalar.  These scalars correspond to the twisted vertex operators, $V_{1,\mathrm{T}}$ and $V_{2,\mathrm{T}}$.  For $V_{1,\mathrm{T}}$ the unnormalized polarization tensor is given by
\be
\eps_{\mathrm{T}}^{MN;\tM\tN}=
\sfrac12\bigbrk{\hat\eta^{M\tM}\hat\eta^{N\tN}+\hat\eta^{M\tN}\hat\eta^{N\tM}}-\sfrac{1}{9}\hat\eta^{MN}\hat\eta^{\tM\tN}\,,
\label{eq:epsmassive}
\ee
where $\hat\eta^{MN}\equiv\eta^{MN}-\frac{k^M k^N}{k^2}$.  For $V_{2 ,\mathrm{T}}$ the unnormalized polarization is
\be
\al_{\mathrm{T}}^{MNL;\tM\tN\tL}=
\sfrac{1}{3!^2}\bigbrk{\hat\eta^{M\tM}\hat\eta^{N\tN}\hat\eta^{L\tL}\mp(5\text{ permutations})}\,.
\label{eq:alphaT}
\ee
We will also need the vertex operators in the $(0,0)$ picture, and these are given by
\begin{align}
V_{1,\mathrm{T},k}^{(0,0)}&=-\gc\left(\stap\right)^2\eps_{\mathrm{T},MN;\tM\tN}\big(i\p X^M(i\p X^N+\sapt k\scdot\psi\,\psi^N)+\sapt\p\psi^M\psi^N\big)\nn\\
&\qquad\qquad\qquad\qquad\times\big(i\bar\p X^\tM(i\bar\p X^\tN+\sapt k\scdot\tilde\psi\,\tilde\psi^\tN)+\sapt\bar\p\tilde\psi^\tM\tilde\psi^\tN\big)e^{ik\cdot X}\,,
\label{eq:V100}
\\
V_{2,\mathrm{T},k}^{(0,0)}&=-\gc\stap\,\al_{\mathrm{T},MNL;\tM\tN\tL}\bigbrk{3\,i\p X^M+\sapt k\scdot\psi\,\psi^M}\psi^N\psi^L
\nn\\&\qquad\qquad\qquad\qquad\times
\bigbrk{3\,i\bar\p X^\tM+\sapt k\scdot\tilde\psi\,\tilde\psi^\tM}\tilde\psi^\tN\tilde\psi^\tL e^{ik\cdot X}\,.
\label{eq:V200}
\end{align}

The vertex operators for level-one R-R string states in the $(-1/2,-1/2)$ picture were shown to have the form~\cite{Koh:1987hm}
\begin{multline}
V_{3,k}^{(-1/2,-1/2)}(z,\bar z)=\gc
\bigbrk{i\bar\p X^M\tilde\Theta-\sfrac18\sapt\tilde\psi^M(\slk\tilde{\slpsi}\tilde\Theta)}^Ae^{-\tilde\phi/2}
\\\times
t_{MA;NB}\bigbrk{i\p X^N\Theta-\sfrac18\sapt\psi^N(\slk\slpsi\Theta)}^Be^{-\phi/2}
e^{ikX}\,,
\label{eq:V3main}
\end{multline}
with the twisted (scalar) states given by~\cite{Minahan:2012fh}%
\footnote{Again we assume that all
spin fields $\tw$, $\tilde\tw$ in \eqref{eq:V3main}
have positive chirality.}
\be
t_{\mathrm{T},MA;NB}=\left(\sapt\right)^{1/2}\bigbrk{C^\dagger\slk(\hat\eta_{MN}-\sfrac19\hat\Gamma_M\hat\Gamma_N)}_{AB}\,,\qquad
\hat\Gamma^M=\Gamma^M-{\slk k^M}/{k^2}\,.
\label{eq:tmassive}
\ee
If we now take a linear combination of $V_{1,\mathrm{T}}$, $V_{2,\mathrm{T}}$ and $V_{3,\mathrm{T}}$ and solve for \eqref{twvc} up to spurious terms, we find that
\be\label{twvert}
V_{\mathrm{T}}=
V_{1,\mathrm{T}}+
V_{2,\mathrm{T}}+
\sfrac{1}{\sqrt{2}}V_{3,\mathrm{T}}
\ee
does the trick up to an overall normalization constant.  This vertex can then be untwisted in a fashion similar to the massless case.  Further technical details including a careful derivation of the supersymmetry transformations and the normalization factors can be found in \appref{massivevert}.

\section{Three-Point Functions}
\label{sec:results}

In this section we compute the three-point functions for any
combination of chiral primaries and level-one primaries, where we use the
properly normalized and untwisted versions of the vertex operators
\eqref{eq:WT,twvert}.

The chiral primaries are operators of the form
\be
\OO\indup{CP}=C_{I_1I_2\dots I_J}\tr[\Phi^{I_1}\Phi^{I_2}\dots\Phi^{I_J}]\,,
\ee
where $C_{I_1I_2\dots I_J}$ is symmetric and traceless.  These are
states in the $[0,J,0]$ representation of $\grp{SO}(6)$ and have the
protected dimension $\Delta\indup{CP}=\Delta_0=J$.

The other operators that make up the three-point functions are assumed to be primaries dual to string states at level one.  These operators also transform in the $[0,J,0]$ representation, but now the bare dimension is $\Delta_0=J+2$~\cite{Beisert:2002tn}. Operators of this type include those with the form
\be
\OO_J=\tr[\Phi^I\Phi_I Z^J]+\dots
\ee
where $\Phi_I$ are the six scalars of \sym,
$Z=\brk{\Phi_5+i\Phi_6}/\sqrt{2}$, and the ellipsis refers to different positions of the $\Phi_I$ in the trace, such that the corresponding magnon momenta are at level one~\cite{Beisert:2002tn,Minahan:2002ve}.  If $J>0$ then levels higher than one are possible, but level one is the only possibility for the Konishi operator
\begin{equation}
\OO\indup{K}=\OO_{J=0}=\tr[\Phi^I\Phi_I]\,.
\end{equation}

Our results are applicable if $\Delta\gg1$.  For the chiral primaries this means that $J\gg1$, but for the level-one primaries small values of $J$ are possible, as long as the $R$-charges in the full three-point function are conserved.

One should note that correlators with two or more twisted operators
must vanish due to a supersymmetric Ward identity.%
\footnote{We thank E. Witten for a discussion on this point.}
The twisted vertices are annihilated by the sixteen supercharges
$Q^A=Q\indup{L}^A-i\,Q\indup{R}^A$ and, since they have non-zero
momenta, they can be written as the action of those supercharges on
another vertex operator $V\indup{T}=\{Q^A,X\}$. It is
then straightforward to see that
\begin{equation}
\bigvev{V\indup{1,T} V\indup{2,T} V\indup{3,T}}
=
\bigvev{X\,\comm{Q^A}{V\indup{2,T}}\,V\indup{3,T}}
+\bigvev{X\,V\indup{2,T}\,\comm{Q^A}{V\indup{3,T}}}
=
0\,.
\end{equation}
If instead we consider the correlator $\vev{V\indup{1,T} V\indup{2,T}
V\indup{3}}$, the same argument follows except that we must choose the
supersymmetries that also annihilate the untwisted vertex $V_3$, which
correspond to the eight supercharges $Q^-_{\alpha\tilde a}=Q\supup{L}_{\al \tilde
a}-i\,Q\supup{R}_{\al \tilde a}$.
In the case $\vev{V\indup{1,T} V\indup{2}
V\indup{3}}$ of two untwisted vertices with momenta $k_{2,3}$, if one of them is closed under
$Q^-_{\alpha\tilde a}$, $Q^{+,\tilde
a}_{\dot{\alpha}}=Q^{\mathrm{L},\tilde
a}_{\dot{\alpha}}+i\,Q^{\mathrm{R},\tilde a}_{\dot{\alpha}}$, then the
other is closed under another set of $Q^{\pm}$, which is obtained from $Q^-_{\alpha\tilde a}$, $Q^{+,\tilde
a}_{\dot{\alpha}}$ by a rotation that takes $k_2$ into $k_3$. The
rotation mixes $\alpha$ with $\dot{\alpha}$ components and lower with
upper $\tilde a$ components, and thus there is no common supercharge
that annihilates both untwisted vertex operators.
Hence, only correlators with at least two untwisted vertices can be
non-zero.

\subsection{Three Chiral Primaries}

In this subsection we compute the three-point function for three chiral primaries, showing that it is consistent with the result in~\cite{Lee:1998bxa}.

As shown in \appref{sec:masslessvert}, the normalized vertex operator satisfying the primary operator condition is given by
\be\label{masslessnormvert}
W_k=-\sfrac{1}{4}\bigbrk{W_{1,k}+\sfrac{1}{\sqrt{2}} W_{2,k}}\,.
\ee
We have included an extra sign that is not determined by the
$Q_{\mathrm{L}}=\pm i\,Q_{\mathrm{R}}$ conditions nor by normalization, but is necessary to be consistent with~\cite{Lee:1998bxa}.  We will continue to use this choice of sign in the later subsections.

The three-point amplitudes are of two general types, with either zero
or two R-R vertex operators inside the amplitude.  For NS-NS vertex
operators only, two of the vertex operators should be given in the
$(-1,-1)$ picture, while one should be in the $(0,0)$ picture to have
the correct ghost charge.  In \appref{sec:masslesscont} it is shown that
\begin{equation}\label{3NSml}
\vev{W_{1,k_1}^{(-1,-1)}W_{1,k_2}^{(-1,-1)}W_{1,k_3}^{(0,0)}}=\frac{4\gc^3\,\ap\tilde\al_1\tilde\al_2\tilde\al_3\tilde\Sigma}{J_1^2J_2^2J_3^2}
\bigbrk{J_1^4+6 J_1^2J_2^2+J_2^4+6J_1^2J_3^2+6J_2^2J_3^2+J_3^4}\,.
\end{equation}
For two R-R vertex operators and one NS-NS, we can choose the R-R
vertex operators to be in the $(-1/2,-1/2)$ picture and the NS-NS in the
$(-1,-1)$ picture.  In \appref{sec:masslesscont} we then show that
\begin{equation}\label{1NSml}
\vev{W_{1,k_1}^{(-1,-1)}W_{2,k_2}^{(-1/2,-1/2)}W_{2,k_3}^{(-1/2,-1/2)}}=\frac{32\gc^3\,\ap\tilde\al_1\tilde\al_2\tilde\al_3\tilde\Sigma}{J_1^2J_2J_3}\bigbrk{3J_1^2+J_2^2+J_3^2}\,.
\end{equation}

Taking the combination in \eqref{masslessnormvert} and using \eqref{3NSml} and \eqref{1NSml}, we find
\be
\label{eq:www}
\vev{W_{k_1}W_{k_2}W_{k_3}}
=
\gc^3\,\ap\,\frac{\tilde\al_1\tilde\al_2\tilde\al_3\tilde\Sigma^5}{J_1^2J_2^2J_3^2}\,,
\ee
with $\tilde\Sigma$ and $\tilde\al_i$ defined in \eqref{tildedefs}.

Using \eqref{3corr2}, \eqref{coupling}, \eqref{3S5}, \eqref{CCC} and
the AdS/CFT dictionary value $\gc=\pi^{3/2}/N$, we obtain
\be
\CC_{123}\approx\frac{1}{N}\sqrt{J_1J_2J_3}\,\vev{C^{J_1}C^{J_2}C^{J_3}}\,,
\label{eq:cwww}
\ee
reproducing the result in~\cite{Lee:1998bxa}.

\subsection{Two Chiral Primaries}

We will
next compute
the correlator for two chiral primaries
and one massive operator $\vev{W_{k_1}W_{k_2}V_{k_3}}$. As shown in
\appref{sec:primaryvertices}, the normalized vertex operators which satisfy the primary condition are \eqref{masslessnormvert} and
\be\label{massivenormvert}
V_k=-\sfrac{1}{16}\bigbrk{V_{1,k}+V_{2,k}+\sfrac{1}{\sqrt 2} V_{3,k}}\,.
\ee
Here the overall minus sign is again not determined with the primary operator condition, but we make this choice to keep the sign of the structure constants obtained consistent with the ones at weak coupling.
Again, we have two types of three-point amplitudes. The first, with no
R-R vertex operators, consists of $\vev{W_{1,k_1}W_{1,k_2}V_{1,k_3}}$ and
$\vev{W_{1,k_1}W_{1,k_2}V_{2,k_3}}$,
where two operators have ghost charges $(-1,-1)$ and one has $(0,0)$.
The second group, with two R-R vertices, is formed by
$\vev{W_{2,k_1}W_{2,k_2}V_{1,k_3}}$,
$\vev{W_{2,k_1}W_{2,k_2}V_{2,k_3}}$ and
$\vev{W_{1,k_1}W_{2,k_2}V_{3,k_3}}$,
with the R-R operators in the $(-1/2-1/2)$ frame and the NS-NS in the $(-1,-1)$.

All these correlators are computed in \appref{sec:contract2chiral}. Using \eqref{masslessnormvert} and \eqref{massivenormvert} we find
\be
\label{eq:wwv}
\vev{W_{k_1}W_{k_2}V_{k_3}}=\gc^3\,\ap^2\,\frac{\al_1^2\al_2^2\Sigma^4\tilde\al_3^2\tilde\Sigma^2}{\Delta_1^2\Delta_2^2\Delta_3^4}\,.
\ee
For the massless states \eqref{masslessnormvert}, we have $\Delta=J$.
When $J_1=J_2\equiv J$ and $J_3=0$, i.e.\
$\Delta\equiv\Delta_3=2/\sqrt{\ap}$, the correlator is given by
\be
\vev{W_{k_1}W_{k_2}V_{k_3}}=\sfrac{\gc^3}{16}\ap^2(J+\half\Delta)^4\,.
\ee
Using \eqref{3corr2} and \eqref{coupling}, the structure constant
becomes
\begin{equation}
\CC_{123}\approx
\frac{\sqrt{\pi}}{4N\sqrt{\lambda}}
\,2^{-\Delta}
J^{2(1-J)}
(J-\half\Delta)^{J-\Delta/2-1/2}
(J+\half\Delta)^{J+\Delta/2+3/2}\,,
\label{eq:cwwv}
\end{equation}
where we used that
$\vev{\psi_{J_1}\psi_{J_2}\psi_{J_3}}=1/\pi^{3/2}$ in this case.

We should say that the
procedure we are using here
requires that the operator dimensions satisfy a triangle rule,
$\Delta_1+\Delta_2\ge\Delta_3$, which means that the result in
\eqref{eq:cwwv} is only valid if $J\ge\Delta/2$.  As we approach the
extremal case $\Delta_1+\Delta_2=\Delta_3$, the saddle point approaches
the boundary~\cite{Klose:2011rm}, and \eqref{3corr2} develops a
singularity~\cite{D'Hoker:1999ea}.  The same thing can happen for
three chiral primaries, but in that case the coupling also goes to
zero and the overall structure constant stays finite~\cite{D'Hoker:1999ea}.
That is not the case here as the coupling is
non-zero at the extremal value.  Furthermore, because of $R$-charge
conservation it is not possible to go beyond the extremal case for
three chiral primaries, but it is possible when one operator is
non-chiral. Lastly, it is known that three-point
functions of single-trace operators get enhanced at the extremal point due to
$1/N$ mixing with double-trace operators~\cite{D'Hoker:1999ea}.
This all suggests a breakdown of the particle
approximation at the extremal point.

\subsection{One Chiral Primary}

For the computation of the correlator $\vev{V_{k_1}V_{k_2}W_{k_3}}$ we use again the normalizations \eqref{masslessnormvert} and \eqref{massivenormvert}.
The amplitudes with no R-R vertex operators are
$\vev{V_{1,k_1}V_{1,k_2}W_{1,k_3}}$,
$\vev{V_{2,k_1}V_{2,k_2}W_{1,k_3}}$ and
$\vev{V_{2,k_1}V_{1,k_2}W_{1,k_3}}$,
where two operators have ghost charges $(-1,-1)$ and one has $(0,0)$.
The second group, with two R-R vertices in the $(-1/2,-1/2)$
picture and the NS-NS vertex in the $(-1,-1)$ picture, is formed by
$\vev{V_{1,k_1}V_{3,k_2}W_{2,k_3}}$,
$\vev{V_{2,k_1}V_{3,k_2}W_{2,k_3}}$ and
$\vev{V_{3,k_1}V_{3,k_2}W_{1,k_3}}$.
Using the formulas for these correlators from
\appref{sec:contract1chiral}, we get in this case a more complicated result
\begin{align}
\vev{V_{k_1}V_{k_2}W_{k_3}}=
\frac{\gc^3\,\Sigma^4}{\Delta_1^4\Delta_2^4\Delta_3^2}
\Bigbrk{
-\half\Omega_1^2\al_3^4+\al_3^2\lrbrk{3\Omega_1^2+2\Omega_1\al_3+\al_3^2}\Omega_4
-\half\Omega_1(\Omega_1-4\al_3)\Omega_4^2+\Omega_4^3
}\,,
\label{eq:vvw}
\end{align}
where
\begin{equation}
\Omega_1=\alpha_1+\alpha_2\,,
\qquad
\Omega_4=\ap\alpha_1\alpha_2\alpha_3\Sigma\,.
\end{equation}
The result simplifies if we assume all $J_i\ll
\la^{1/4}$, in which case we find
\be
\vev{V_{k_1}V_{k_2}W_{k_3}}\approx\sfrac{\gc^3}{2}\bigbrk{1+2\sqrt{\ap} J_3+\ldots}\,.
\ee
Using \eqref{3corr2}, \eqref{coupling} we obtain in this limit that the structure constant is
\be
\CC_{123}
\approx
\frac{\pi^2}{N}
\,2^{-J_3}
\bigbrk{\lambda^{1/4}+\sfrac{3}{2}J_3+\order{\lambda^0}}
\,\vev{\psi_{J_1}\psi_{J_2}\psi_{J_3}}\,,
\label{eq:cwvv}
\ee
Choosing $J_1=0$ and $J_2=J_3=J$, it becomes
\begin{equation}
\CC_{123}\approx\,\frac{1}{N}\,\pi^{1/2}\,(\lambda^{1/4}+\sfrac{3}{2}J)\,2^{-J}\,.
\end{equation}

\subsection{Zero Chiral Primaries}

Finally, we compute the three-point function of three Konishi-like
operators $\vev{V_{k_1}V_{k_2}V_{k_3}}$, using the vertex \eqref{massivenormvert}.
The amplitudes with no R-R vertex operators are
$\vev{V_{1,k_1}V_{1,k_2}V_{1,k_3}}$,
$\vev{V_{2,k_1}V_{2,k_2}V_{2,k_3}}$,
$\vev{V_{2,k_1}V_{2,k_2}V_{1,k_3}}$ and
$\vev{V_{1,k_1}V_{1,k_2}V_{2,k_3}}$,
where two operators have ghost charges $(-1,-1)$ and one has ghost
charge $(0,0)$.
The correlators with two R-R vertices in the $(-1/2,-1/2)$
picture and one NS-NS vertex in the $(-1,-1)$ picture are
$\vev{V_{1,k_1}V_{3,k_2}V_{3,k_3}}$ and
$\vev{V_{2,k_1}V_{3,k_2}V_{3,k_3}}$.
Once again, the details of the calculation of these amplitudes are
given in \appref{sec:contract0chirals}. Putting all cases together we get
\begin{multline}
\vev{V_{k_1}V_{k_2}V_{k_3}}=
\frac{\gc^3\,\Sigma^4}{\Delta_1^4\Delta_2^4\Delta_3^4}
\Bigbrk{
\half\Sigma_2^4+\sfrac{9}{2}\Sigma_4^2+(2\Sigma_2^3-3\Sigma^2\Sigma_2^2+6\Sigma^2\Sigma_4+3\Sigma_2\Sigma_4)\ap\Sigma_4
\\
+\half(3\Sigma^4+7\Sigma_2^2-8\Sigma^2\Sigma_2+6\Sigma_4) (\ap\Sigma_4)^2
-(\Sigma^2-3\Sigma_2)(\ap\Sigma_4)^3+(\ap\Sigma_4)^4
}\,.
\label{eq:vvv}
\end{multline}
where
\begin{equation}
\Sigma_2=\alpha_1\alpha_2+\alpha_1\alpha_3+\alpha_2\alpha_3\,,
\qquad
\Sigma_4=\alpha_1\alpha_2\alpha_3\Sigma\,,
\end{equation}
If we now consider the case of all $J_i=0$ (i.e.\
$\Delta_i=2/\sqrt{\ap}$), we obtain the three-point function \eqref{eq:VVVexp}
for three Konishi operators
\be
\vev{V_{k_1}V_{k_2}V_{k_3}}\approx\gc^3\frac{3^8}{2^9}\,.
\ee
Plugging this result in \eqref{3corr2}, we find the structure constant in
\eqref{CCresulta},
\begin{equation}
\CC_{123}\approx\frac{1}{N}\,64\,{\pi^{1/2}}\lambda^{1/4}\lrbrk{\frac{3}{4}}^{3\lambda^{1/4}+5/2}\,.
\end{equation}

\section{Discussion}
\label{sec:conclusions}

In this work we computed the three-point functions for certain short scalar
operators in \sym\ at strong coupling. The set of
operators we considered are the chiral primaries as well as primaries which are dual to string states at the first massive level, which includes the Konishi operator.
Following the analysis
in~\cite{Janik:2010gc,Minahan:2012fh,Klose:2011rm},
the action is dominated by a saddle point
composed of three geodesic trajectories propagating from the AdS
boundary to a common intersection point, whose location is determined
by the conservation of the canonical momentum. For short operators
(whose dimensions scale as $\Delta\sim\sqrt[4]{\lambda}$), the
interaction region is small compared to the AdS
radius, and hence the coupling at the intersection point is
well approximated
by a
flat-space correlator of appropriate type IIB string vertex operators. For
primary operators, the vertex operators have to satisfy a twisted version of
the $Q\indup{L}=i \,Q\indup{R}$ equation. This constraint uniquely singles
out the appropriate string states at the massless and first massive
level.%
\footnote{There is only one long multiplet at the first massive level,
and hence a unique solution to $Q\indup{L}=i\,Q\indup{R}$.}
The corresponding vertex operators are certain twisted linear
combinations of ten-dimensional flat-space NS-NS and R-R scalar modes.
In addition, there is a contribution from the wave-function overlap on
the~$\grp{S}^5$ part of the space.

Identifying the correct string vertex operators corresponding to given
local operators in the boundary CFT is a long-standing open problem in
the AdS/CFT duality. The procedure applied here
could provide insights
as to what
the full $\grp{AdS}_5\times\grp{S}^5$ vertex operators
might
look like. It should
also
be mentioned that the general method used here and in~\cite{Minahan:2012fh} is not
just
restricted to \sym\
with its IIB dual
in $\grp{AdS}_5\times\grp{S}^5$. It
should in principle apply to all AdS/CFT theories
with a string theory as a gravitational dual,
up to the
construction of the appropriate vertex operators. Here we have relied
on $\superN=4$ supersymmetry to identify the vertex operators.

Our results for the structure constants for any combination of
scalar
massless and
level-one
states are given in
\eqref{CCresult}, \eqref{eq:cwww}, \eqref{eq:cwwv}, and \eqref{eq:cwvv}
above. Carrying out the contractions of the primary vertices requires
substantial effort due to the
combinatorial complexity.
Moreover, many intermediate results (partial correlators)
contain unappealing  prime factors.
In comparison, the complete three-point
correlators are astonishingly simple. Of course this is not completely
unexpected, as planar \sym\ is believed to be integrable.%
\footnote{See~\cite{Beisert:2010jr} for an introductory review.}
The simplicity of our results suggests that integrability will play an
equally important role in the computation of three-point functions as
it did for the spectral problem.
But even disregarding integrability, our results demonstrate that the
holographic computation of massive correlators, albeit technically
involved, are in fact feasible.

We find that the structure constants for massive states typically are
exponentially small in the scaling dimension
$\Delta\sim\sqrt[4]{\lambda}$. However, the suppression is still
moderate in interesting regimes. For example, the structure constant
\eqref{CCresulta} for three Konishi states is
$\mathcal{C}_{123}\approx2.4/N$ for $\lambda=10^3$, a regime in which
the Konishi dimension clearly shows GKP behavior~\cite{Gromov:2009zb} and hence the
semiclassical approximation should be applicable.
Also, in the case of one chiral primary and two massive primaries, the
suppression is only exponential in the R-charge $J$ of the chiral primary
operator.

\bigskip
\noindent
There are several directions for future work:
It would be very interesting to compute subleading corrections to our
results in the $\ap$ expansion.
Subleading terms can
be obtained by inserting the expansions for
the dimensions $\Delta_{1,2,3}$ of the respective operators, and by
calculating loop corrections to the worldsheet amplitude of the appropriate
untwisted vertex operators.
Analytic corrections to our results will be an expansion in powers of
$\ap$. There is a possibility of non-analytic corrections that come
with half-integer powers, as happens for the operator
dimensions~\cite{Roiban:2011fe}. But we suspect that such terms only
enter the
correlators indirectly, through the corrections to the operator
dimensions.
The three-point vertices are functions of the momenta, whose
corrections will come in half-integer powers of $\ap$ through their
dependence on the scaling dimensions.
It is also likely that  \eqref{3corr2} itself will get modified as we move away from the point-like limit, as  indicated by the singularities that appear as we approach the extremal limit.

Another exciting venture would be to study four-point functions of
short operators, aiming at a holographic operator product expansion.
For related references, see
e.g.~\cite{D'Hoker:1999pj,D'Hoker:1999ni,Penedones:2010ue,Fitzpatrick:2011ia,Costa:2012cb}.
It can be hoped that the methods used here will also be useful for
higher-point correlators.  Likewise, information from the four-point
correlation functions of protected operators can be used to check the
validity of our results.  For example, in~\cite{Costa:2012cb} there is
a prediction for the structure constant between two operators dual to
the dilaton and a third operator dual to a higher-spin state on the
graviton Regge trajectory.  It would  be interesting to find
their result for the spin~4 operator, which is at the first massive
level, using the methods here.

Finally, a bigger goal would be to employ integrability in the
computation of short-string correlators. Based on the experience that
integrability techniques usually start with a large-volume limit, this
would probably require the adaption of a full thermodynamic Bethe
ansatz~\cite{Arutyunov:2007tc,Arutyunov:2009ur,Gromov:2009tv,Bombardelli:2009ns,Gromov:2011cx}
or quantum spectral curve~\cite{Gromov:2013pga} machinery.

\subsection*{Acknowledgments}

We thank M.~Bianchi, F.~Morales, J.~Penedones, O.~Schlotterer and
E.~Witten for discussions.
This research is supported in part by
Vetenskapsr{\aa}det under grant \#2012-3269.
The work of TB is supported by a Marie Curie International Outgoing
Fellowship
within the $7^{\mathrm{th}}$ European Community Framework Programme
under grant PIOF-GA-2011-299865.
JAM thanks the CTP at MIT and the Galileo Galilei Institute for kind
hospitality during the course of this work.

\appendix

\addtocontents{toc}{\protect\setcounter{tocdepth}{1}}

\section{Bosonization Setup and OPE's}
\label{sec:setup}

We follow the conventions of Polchinski~\cite{Polchinski:1998stringboth}.
In all products of operators evaluated at the same worldsheet point,
normal ordering is implicitly assumed.

\paragraph{Bosonization.}

For OPE's and correlators involving R-R vertices, we use bosonization
as described in~\cite{Friedan:1985ge,Kostelecky:1986xg}. Namely, the
ten real fermions $\psi^M$ are written as five complex bosons
$\phi^j$. First convert the $\psi^M$ to a Cartan--Weyl basis,
\begin{equation}
\psi^{\pm e_0}=\frac{1}{\sqrt{2}}\Bigbrk{\pm\psi^{0}+\psi^{1}}\,,
\qquad
\psi^{\pm e_j}=\frac{1}{\sqrt{2}}\Bigbrk{\psi^{2j}\pm i\psi^{2j+1}}\,,
\qquad
j=1,\dots,4\,.
\label{eq:cwbasis}
\end{equation}
Here, $e_j$ denotes a unit vector in direction $j$ on the Euclidean weight lattice
of $\alg{o}(1,9)$, where $\pm e_i\pm e_j$ form the non-zero roots.
Now write the $\psi^{\pm e_j}$ as
\begin{equation}
\psi^{\pm e_j}(z)=e^{\pm e_j\cdot\phi(z)}c_{\pm e_j}\,,
\qquad
c_{\pm e_j}=(-1)^{N_1+\ldots+N_{j-1}}=e^{i\pi\brk{\pm e_j\cdot M\cdot N}}\,,
\end{equation}
where $\phi=(\phi^1,\dots,\phi^5)$ is the vector of bosonic fields,
and $c_{\pm e_j}$ is a ``cocycle factor'' that ensures the right
commutation relations: $N=\brk{N_1,\dots,N_5}$ is the vector of
fermion number operators $N_j$ for the fermions $\psi^{\pm e_j}$, and
$M$ is a lower-triangular sign matrix, given in Appendix~E
of~\cite{Kostelecky:1986xg}. More generally, any operator of conformal
weight $\lambda$ can be written as $e^{\lambda\cdot\phi}c_{\lambda}$.
In particular, the twist fields $\tw^A$ become
\begin{equation}
\tw^A=e^{A\cdot\phi}c_A\,,
\qquad
c_A=e^{i\pi A\cdot M\cdot N}\,,
\label{eq:spinfields}
\end{equation}
where $A=(\pm,\dots,\pm)/2$ is a five-component spinor weight. Now the
OPE for two bosonized operators of weight $\lambda$, $\lambda'$ takes
the form
\begin{align}
e^{\lambda\cdot\phi(z_1)}c_{\lambda}e^{\lambda'\cdot\phi(z_2)}c_{\lambda'}
&=
z_{12}^{\lambda\cdot\lambda'}e^{i\pi\lambda\cdot M\cdot\lambda'}
\normord{e^{\lambda\cdot\phi(z_1)+\lambda'\cdot\phi(z_2)}}\,c_{\lambda+\lambda'}
\nn\\
&=
z_{12}^{\lambda\cdot\lambda'}e^{i\pi\lambda\cdot M\cdot\lambda'}
\normord{e^{(\lambda+\lambda')\cdot\phi(z_2)}
\Bigbrk{1+z_{12}\,\lambda\scdot\p\phi(z_2)+\order{z_{12}}^2}}\,c_{\lambda+\lambda'}
\label{eq:bosonOPE}
\end{align}
Including the bosonic ghost field $\phi\indup{g}\equiv\phi^6$, the
weight lattice becomes Lorentzian
with a ``timelike'' sixth dimension, and with the sign matrix $M$ now
given by (E.7) in~\cite{Kostelecky:1986xg}. The OPE formula
\eqref{eq:bosonOPE} continues to hold.
For example,
\begin{equation}
\psi^{\pm e_j}e^{-\phi\indup{g}}
=
e^{\lambda\cdot\phi}c_{\lambda}\,,
\qquad
\lambda=\pm e_j+e_6\,,
\qquad
c_{\lambda}
=
e^{i\pi\lambda\cdot M\cdot N}\,.
\end{equation}
Using the OPE \eqref{eq:bosonOPE}, the correlator of $n$ fields with
(six-dimensional) weights $\lambda_1,\dots,\lambda_n$ becomes
\begin{equation}
\bigvev{e^{\lambda_1\cdot\phi(z_1)}c_{\lambda_1}\dots
e^{\lambda_n\cdot\phi(z_n)}c_{\lambda_n}}
=
\delta\bigbrk{{\textstyle\sum_j\lambda_j-2e_6}}
\prod_{1\leq j<k\leq n}
z_{jk}^{\lambda_j\cdot\lambda_k}
e^{i\pi\lambda_j\cdot M\cdot\lambda_k}\,,
\end{equation}
where the scalar product in the exponents is Lorentzian with signature
$(1,\dots,1,-1)$.

\paragraph{Gamma Matrices.}

In the Cartan--Weyl basis, we use the gamma matrices
\begin{equation}
\bigbrk{\Gamma^{\pm e_k}}^A{}_B=\sqrt{2}\,e^{i\pi(\pm e_k)\cdot M\cdot A}\delta(\pm e_k+A,B)\,,
\qquad
k=1,\dots,5\,,
\qquad
A=(\pm,\dots,\pm)/2\,.
\end{equation}
They are related to the gamma matrices $\Gamma^M$ in the covariant
basis in the same way $\psi^{\pm e_j}$ are related to $\psi^M$
\eqref{eq:cwbasis}.
For the charge conjugation matrix we use
\begin{equation}
\begin{split}
C^{AB}
&=e^{i\pi 3/4}\sigma^2\otimes\sigma^1\otimes\sigma^2\otimes\sigma^1\otimes\sigma^2\\
&=\delta(A+B)
\begin{cases}
 e^{-i\pi A_{-1/2}\cdot M\cdot A_{-1/2}} & \text{for $A$ with positive chirality}\\
-e^{-i\pi A_{+1/2}\cdot M\cdot A_{+1/2}} & \text{for $A$ with negative chirality}
\end{cases}
\end{split}
\end{equation}
Here, $\sigma^j$ are the usual Pauli matrices, and
$A_{\pm1/2}=(A,\pm\half)$ is the spinor weight $A$ extended by the ghost weight
$\pm 1/2$.

\paragraph{OPE's.}

The non-vanishing OPE's among
the elementary fields are
\begin{equation}
X_i^MX_j^N
=
-\apt\eta^{MN}\log\abs{z_{ij}}^2+\normord{X_i^MX_j^N}\,,
\qquad
\psi_i^M\psi_j^M
=
\frac{\eta^{MN}}{z_{ij}}+\normord{\psi_i^M\psi_j^N}\,,
\end{equation}
where subscript $_i$ denotes dependence on worldsheet coordinates
$(z_i,\bar z_i)$, and $z_{ij}\equiv z_i-z_j$.
In three-point correlators, one often needs to use the OPE of a
coordinate derivative with two plane-wave factors, which takes the
convenient form
\begin{equation}
i\p X_1^Me^{ik_2\cdot X_2}e^{ik_3\cdot X_3}
=
\apt\frac{k_2z_{23}}{z_{12}z_{13}}e^{ik_2\cdot X_2}e^{ik_3\cdot X_3}
+\dots
\qquad
\text{for }k_1+k_2+k_3=0\,.
\end{equation}
For the correlator of plane-wave factors, we use
\begin{equation}
\Bigvev{\prod_{j=1}^n\normord{e^{ik_j\cdot X_j}}}
=
\delta\bigbrk{{\textstyle\sum_jk_j}}\prod_{1\leq j<k\leq n}\abs{z_{jk}}^{\ap k_j\cdot k_k}\,.
\end{equation}
where for simplicity of notation we write
$\delta\bigbrk{\sum_jk_j}=(2\pi)^{10} \delta^{10}\bigbrk{\sum_jk_j}$.
The OPE's for bosonized fields are given in \eqref{eq:bosonOPE}.
Further relevant OPE's for bosonized fields~$\phi^j$ are ($j=1,\dots,6$)
\begin{align}
\phi^i(z_1)\phi^j(z_2)
&=
\gamma^{ij}\log\abs{z_{12}}^2+\normord{\phi^i(z_1)\phi^j(z_2)}\,,
\qquad
\gamma=\diag(1,1,1,1,1,-1)\,,
\\
\p\phi^j(z_1)e^{\lambda\cdot\phi(z_2)}
&=
\frac{\lambda\cdot
e_j}{z_{12}}e^{\lambda\cdot\phi(z_2)}+\normord{\p\phi^j(z_1)e^{\lambda\cdot\phi(z_2)}}
\end{align}
When computing three-point functions, the worldsheet integrations will
be replaced by the corresponding $c$, $\tilde c$ ghosts. Three-point
correlators are then accompanied by the following expectation value
\begin{equation}
\bigvev{c_1\,c_2\,c_3\,\tilde c_1\,\tilde c_2\,\tilde c_3} =|z_{12}|^2|z_{13}|^2|z_{23}|^2
\end{equation}
For simplicity of notation, in the following sections we will omit
these ghosts from the correlators and their presence will be implicit,
making the correlators have no dependence on the worldsheet
coordinates.

\paragraph{Point-Splitting.}

We define products of fields evaluated at the same worldsheet
coordinate by a point-splitting regularization, see
e.g.~\cite{Koh:1987hm}. Specifically,
\begin{gather}
\psi^{+e_i}\psi^{+e_i}(z)
=\psi^{-e_i}\psi^{-e_i}(z)
=0\,,\nn
\\
\psi^{+e_i}\psi^{-e_i}(z)
=-\psi^{-e_i}\psi^{+e_i}(z)
=\p\phi^i(z)\,.
\label{eq:psipsisplit}
\end{gather}
Moreover, for $i\neq j$ (the two signs $\pm$ are independent),
\begin{multline}
\psi^{\pm e_i}\psi^{\pm e_j}\tw^A(z)
=e^{i\pi(\pm e_i)\cdot M\cdot(\pm e_j)}
e^{i\pi(\pm e_i\pm e_j)\cdot M\cdot A}
\cdot\\\cdot
\begin{cases}
(\pm e_i\pm e_j)\cdot\p\phi\,\tw^{A\pm e_i\pm e_j}(z) & \text{for }(\pm e_i\pm e_j)\scdot A=-1\,,\\
\bfm{\tw}^{A\pm e_i\pm e_j}(z) & \text{for }(\pm e_i\pm e_j)\scdot A=0\,,\\
0 & \text{for }(\pm e_i\pm e_j)\scdot A=1\,.
\end{cases}
\end{multline}
Here, the boldface $\bfm{\tw}^A$ is defined as in \eqref{eq:spinfields},
but also for non-spinorial weight $A$.
Applying the point-splitting
prescription to the term $\psi^{\pm e_k}\bigbrk{\slpsi\tw}^A$
appearing in the vertex \eqref{eq:V3main} yields
\begin{equation}
\psi^{\pm e_k}\bigbrk{\slpsi\tw}^A
=
\delta(A_k\mp\half)4\sqrt{2}\,
e^{\pm i\pi e_k\cdot M\cdot A}\bfm{\tw}^{A\pm e_k}
+
\bigbrk{2A\pm6e_k}\cdot\p\phi\,\bigbrk{\Gamma^{\pm
e_k}\tw}^A\,.
\label{eq:psipsiS}
\end{equation}
This also implies
\begin{equation}
\brk{\slpsi\slpsi\tw}^A=36\brk{A\scdot\p\phi}\tw^A\,.
\label{eq:36A}
\end{equation}
When computing correlation functions, terms like \eqref{eq:psipsiS} will
always appear multiplied by the polarization tensor $t_{\tilde M\tilde
A,M A}$ which obeys $t_{\tilde M\tilde A, M A}
(\Gamma^M)^A_{\;\;C}=0$, so we are free to add terms of the form
$B\cdot\p\phi\, (\Gamma^{\pm e_k})^A_{\;\;B}\tw^B$ to
\eqref{eq:psipsiS}. We will then use the following formula
\begin{equation}
\psi^{\pm e_k}\bigbrk{\slpsi\tw}^A
\simeq
\delta(A_k\mp\half)4\sqrt{2}\,
e^{\pm i\pi e_k\cdot M\cdot A}\bfm{\tw}^{A\pm e_k}
-
\sfrac{8}{5}\bigbrk{A\mp\sfrac{3}{2}e_k}\cdot\p\phi\,\bigbrk{\Gamma^{\pm
e_k}\tw}^A\,,
\label{eq:psipsiScorr}
\end{equation}
when computing three-point functions. This will ensure a manifestly
correct dependence on the worldsheet coordinates, since
$(A\mp\sfrac{3}{2}e_k)\cdot (A\pm e_k)=0$.

\section{Primary Vertex Operators}
\label{sec:primaryvertices}

To find the massless and massive primary operators, one needs to
compute the action of the supersymmetries on the flat-space vertices.
We will act with the left supercharge in the $(+1/2)$ picture on the
NS-NS vertices, and compare to the action of the right supercharge in
the $(-1/2)$ picture on the R-R vertices. The supercharges are given by~\cite{Koh:1987hm}
\begin{equation}
\label{eq:QR}
Q^{A (-1/2)}_{\mathrm{R}}=\oint\frac{d\bar z}{2\pi i}\tilde\Theta^Ae^{-\tilde\phi_6 /2}
\end{equation}
and, using the definition of $Q\indup{BRST}$ from~\cite{Polchinski:1998stringboth}, we can
change to the frame with ghost charge $1/2$ in the following way%
\footnote{We will need this operator for $A$ with positive chirality.
For $A$ with negative chirality, \eqref{eq:QL} would have the opposite
sign. Note also that the $i$ in front of $Q\indup{BRST}$ is different from the convention in~\cite{Friedan:1985ge}, but necessary to get the overall sign of the vertex operators in the $(0,0)$ frame consistent with~\cite{Polchinski:1998stringboth}.}
\begin{equation}
\label{eq:QL}
Q^{A (+1/2)}_{\mathrm{L}}
=
\Bigcomm{i\,Q\indup{BRST}}{\,\xi\,Q^{A (-1/2)}_{\mathrm{L}}}
=
\sfrac{1}{\sqrt 2}
\lrbrk{\sfrac{2}{\alphap}}^{1/2}\oint\frac{dz}{2\pi i}i\p
X^M\bigbrk{\G_M\Theta}^Ae^{+\phi_6/2}\,.
\end{equation}

\subsection{Massless Vertex Operators}\label{sec:masslessvert}

The supercharges act on the massless vertices in the following way
\begin{equation}
\label{eq:QW1}
\bigcomm{Q^{A (+1/2)}_{\mathrm{L}}}{W_{1,\mathrm{T}}^{(-1,-1)}}=i\sfrac{\gc}{2}\lrbrk{\sfrac{\alphap}{2}}^{1/2}\eps_{\mathrm{T},M N}\lrbrk{\slk\G^M }^{A}_{\;\;B}\Theta^B e^{-\phi_6/2}\tilde\psi^N e^{-\tilde\phi_6} e^{ik\cdot X}
\end{equation}
and
\begin{equation}
\bigcomm{i\,Q^{A (-1/2)}_{\mathrm{R}}}{W_{2,\mathrm{T}}^{(-1/2,-1/2)}}=-i\sfrac{\gc}{\sqrt 2}\lrbrk{\sfrac{\alphap}{2}}^{1/2}\lrbrk{\G_M\slk }^{A}_{\;\;B}\Theta^Be^{-\phi_6/2}\tilde\psi^Me^{-\tilde\phi_6} e^{ik\cdot X}\,.
\end{equation}
Since $\slk\slk=0$ and
\begin{equation}
k_M\tilde\psi^M e^{-\tilde\phi_6} e^{ik\cdot X}
\end{equation}
is a spurious term, then we can write
\begin{equation}
\label{eq:QW2}
\bigcomm{i\,Q^{A (-1/2)}_{\mathrm{R}}}{W_{2,\mathrm{T}}^{(-1/2,-1/2)}}=i\sfrac{\gc}{\sqrt 2}\lrbrk{\sfrac{\alphap}{2}}^{1/2}\eps_{\mathrm{T},M N}\lrbrk{\slk\G^M }^{A}_{\;\;B}\Theta^Be^{-\phi_6/2}\tilde\psi^Ne^{-\tilde\phi_6} e^{ik\cdot X}\,.
\end{equation}
We then get that the massless vertex operator must be of the form $W_{\mathrm{T}}\sim W_{1,{\mathrm{T}}}+\sfrac{1}{\sqrt 2} W_{2,{\mathrm{T}}}$. Its normalization is
\begin{equation}
\label{eq:NormalizationW}
\frac{1}{\gc^2}
\Bigvev{\bigbrk{W_{1,{\mathrm{T}}}^k+\sfrac{1}{\sqrt 2} W_{2,{\mathrm{T}}}^k}^\dagger\bigbrk{W_{1,{\mathrm{T}}}^{k}+\sfrac{1}{\sqrt 2} W_{2,{\mathrm{T}}}^{k}} } =8+8=16\,.
\end{equation}
Note that the vertex operator $\bigbrk{W_{1,{\mathrm{T}}}^k+\sfrac{1}{\sqrt 2} W_{2,{\mathrm{T}}}^k}^\dagger$ satisfies the opposite condition $Q\indup{L}^A=-i\, Q\indup{R}^A$.
The normalized massless primary state is then given by
\begin{equation}
\label{eq:TwistedW}
W_{\mathrm{T}}=\frac{1}{4}\biggbrk{W_{1,{\mathrm{T}}}+\frac{W_{2,{\mathrm{T}}}}{\sqrt 2}}\,.
\end{equation}

\subsection{Massive Vertex Operators}\label{massivevert}

The calculation for the massive case is quite lengthy and tedious so we will present only guidelines of the computation. Starting with $V_{1,{\mathrm{T}}}$ and using that $\eps_{\mathrm{T}, MN,\tilde M\tilde N}$ is symmetric, traceless and orthogonal to $k$, one can obtain
\begin{multline}
\bigcomm{Q^{A (+1/2)}_{\mathrm{L}}}{V_{1,{\mathrm{T}}}^{(-1,-1)}}=i\sfrac{\gc}{2}\lrbrk{\sfrac{2}{\alphap}}^{1/2}\bigbrk{\slk\hG^M}^{A}_{\;\;B}\lrbrk{i\p X_N\Theta-\sfrac{1}{8}\sfrac{\alphap}{2}\psi_N\lrbrk{\slk\slpsi\Theta}}^B e^{-\phi_6/2}\hE^{N P}
\\
\times\lrbrk{\sfrac{1}{2}\tilde\psi_{\{M}i\bar\p X_{P\}}-\sfrac{1}{9}\hE_{M P}\brk{\tilde\psi\cdot i\bar\p X}} e^{-\tilde\phi_6} e^{ik\cdot X}\,.
\label{eq:QV1}
\end{multline}
Since we can add total derivatives to the vertices, one can use the following equation to eliminate terms with $\p\phi_6$
\begin{equation}
\half\p\phi_6\Theta^B e^{-\phi_6/2}e^{ik\cdot X}=i k\scdot\p X\Theta^B e^{-\phi_6/2}e^{ik\cdot X}+\sfrac{1}{36}\lrbrk{\slashed\psi\slashed\psi\Theta}^B e^{-\phi_6/2}e^{ik\cdot X}\,.
\end{equation}
Using that $\alpha_{\mathrm{T},MNP,\tilde M\tilde N\tilde P}$ is antisymmetric and noting that
\begin{align}
&\lrbrk{\sfrac{1}{48}\lrbrk{\G_L+\sfrac{6}{k^2}\slashed k k_L }^{AB}\psi^L\lrbrk{\slashed\psi\Theta}^B+\lrbrk{k_L-\sfrac{1}{4}\slashed k\G_L}^{AB} i\p X^L\Theta^B}e^{-\phi_6/2} e^{ik\cdot X}
\end{align}
is a spurious term, then we obtain that the action of $Q\indup{L}$ on $V_{2,{\mathrm{T}}}$ is
\begin{multline}
\bigcomm{Q^{A (+1/2)}_{\mathrm{L}}}{V_{2,{\mathrm{T}}}^{(-1,-1)}}=i\sfrac{\gc}{4}\lrbrk{\sfrac{2}{\alphap}}^{1/2}\lrbrk{\hG_N\hG_P\bigbrk{\hE_{MR}-\sfrac{1}{9}\hG_M\hG_R}}^{A}_{\;\;B}
\\
\times\lrbrk{i\p X^R\Theta-\sfrac{1}{8}(\sfrac{\alphap}{2})\psi^R\lrbrk{\slashed k\slashed\psi\Theta}}^B e^{-\phi_6/2}\,\tilde\psi^M\tilde\psi^N\tilde\psi^P e^{-\tilde\phi_6} e^{ik\cdot X}\,.
\label{eq:QV2}
\end{multline}
Finally we need only to compute the action of $i\,Q\indup{R}$ on $V_{3,{\mathrm{T}}}$. Using that $t_{\mathrm{T},\tilde M A, M B}$ is orthogonal to $k$ and $(\G^M\, C\, t_{\mathrm{T},M,N})=0$, eliminating terms with $\p\phi_6$ using
\begin{equation}
\bar\p\tilde\phi_6\tilde\psi^M e^{-\tilde\phi_6} e^{ik\cdot X}=i k\scdot\bar\p X\tilde\psi^M e^{-\tilde\phi_6} e^{ik\cdot X}+\bar\p\tilde\psi^M e^{-\tilde\phi_6} e^{ik\cdot X}\,,
\end{equation}
and noting that the following are spurious terms~\cite{Koh:1987hm}
\begin{gather}
\lrbrk{\stap\sfrac{1}{2}i\bar\p X^{ [M }\tilde\psi^{P]}-\half\tilde\psi^M\tilde\psi^R\tilde\psi^P k_{R}}\bigbrk{\hG_{P}C t_{\mathrm{T},MA;NB}}e^{-\tilde\phi_6}e^{ik\cdot X}\,,
\\
\lrbrk{\bar\p\tilde\psi^M+\sfrac{1}{2} i\bar\p X^{\{M}\tilde\psi^{P\}}k_P}\lrbrk{\slk Ct_{\mathrm{T},MA;NB}} e^{-\tilde\phi_6}e^{ik\cdot X}\,,
\end{gather}
we obtain that
\begin{multline}
\bigcomm{i\,Q^{A (-1/2)}_{\mathrm{R}}}{V_{3,{\mathrm{T}}}^{(-1/2,-1/2)}}=
\\
\lrbrk{\brk{\sfrac{i}{2}\bar\p X^{\{M}\tilde\psi^{P\}}\hG_P+\sfrac{1}{4}\lrbrk{\sfrac{\alphap}{2}}\tilde\psi^M\tilde\psi^P\tilde\psi^R\hG_P\hG_R\slk}\slk\brk{\hE_{M N}-\sfrac{1}{9}\hG_M\hG_N}}^A_{\;B} e^{-\tilde\phi_6}
\\
\times\lrbrk{-i\sfrac{\gc}{\sqrt 2}}\lrbrk{\sfrac{2}{\alphap}}^{1/2}\lrbrk{i\p X^N\Theta-\sfrac{1}{8}\lrbrk{\sfrac{\alphap}{2}}\psi^N\lrbrk{\slk\slashed\psi\Theta}}^Be^{-\phi_6/2} e^{ik\cdot X}\,.
\label{eq:QV3}
\end{multline}
We can now easily see from equations \eqref{eq:QV1,eq:QV2,eq:QV3} that the primary constraint is satisfied for a vertex operator of the form $V_{\mathrm{T}} \sim V_{1,{\mathrm{T}}}+V_{2,{\mathrm{T}}}+\sfrac{1}{\sqrt 2} V_{3,{\mathrm{T}}}$.
To normalize it we only need to compute the two-point function with its adjoint, which gives
\begin{equation}
\label{eq:Normalization}
\frac{1}{\gc^2}
\Bigvev{\lrbrk{V_{1,{\mathrm{T}}}^k+V_{2,{\mathrm{T}}}^k+\sfrac{1}{\sqrt 2} V_{3,{\mathrm{T}}}^k}^\dagger\lrbrk{V_{1,{\mathrm{T}}}^{k}+V_{2,{\mathrm{T}}}^{k}+\sfrac{1}{\sqrt 2} V_{3,{\mathrm{T}}}^{k}} }=44+84+128=256\,.
\end{equation}
So we have finally obtained that the primary vertex for the first massive level is
\begin{equation}
\label{eq:TwistedV}
V_{\mathrm{T}}=\frac{1}{16}\lrbrk{V_{1,{\mathrm{T}}}+V_{2,{\mathrm{T}}}+\frac{V_{3,{\mathrm{T}}}}{\sqrt 2}}\,.
\end{equation}

\subsection{Untwisting}

The primary constraint \eqref{twvc} was for the twisted vertex operators $W_{\mathrm{T}}$ and $V_{\mathrm{T}}$, but we must compute the three-point function with the untwisted versions of these operators, $W$ and $V$.
For the massless vertex the modifications are
\begin{align}
\eps^{k_i}_{M,\tilde M}&=(-1)^{\sigma_{k_i}(\tilde M)}\eps^{k_i}_{\mathrm{T},M,\tilde M} \,,
\\
t^{k_i}_{A,B}&=\lrbrk{\sfrac{\alphap}{2}}^{1/2}\lrbrk{\op C^\dagger  i\G^{0'}\G^{1'}\G^{2'}\G^{3'} \slk}_{AB} \,.
\end{align}
For the massive case we have the following modification of the polarization tensors
\begin{align}
\eps^{k_i}_{MN,\tilde M\tilde N}&=(-1)^{\sigma_{k_i}(\tilde M)+\sigma_{k_i}(\tilde N)}\eps^{k_i}_{\mathrm{T},MN,\tilde M\tilde N} \,,
\\
\alpha^{k_i}_{MNP,\tilde M\tilde N\tilde P}&=(-1)^{\sigma_{k_i}(\tilde M)+\sigma_{k_i}(\tilde N)+\sigma_{k_i}(\tilde P)}\alpha^{k_i}_{\mathrm{T},MNP,\tilde M\tilde N\tilde P} \,,
\\
t^{k_i}_{\tilde M A,M B}&=(-1)^{\sigma_{k_i}(\tilde M)}\lrbrk{\sfrac{\alphap}{2}}^{1/2}\lrbrk{\op C^\dagger  i\G^{0'}\G^{1'}\G^{2'}\G^{3'}  \slk\lrbrk{\hE^{M\tilde M}-\sfrac{1}{9}\hG^{\tilde M}\hG^{M}}}_{AB} \,.
\end{align}
Here we defined the twist factor
\begin{equation}
\sigma_{k_i}(M)=
\begin{cases}
1 &\text{if $M=0',\ldots, 3'$,}
\\
0 &\text{if $M=4', 5,\ldots, 9$,}
\end{cases}
\end{equation}
and $0',\ldots,3'$ represent the AdS directions perpendicular to $k_i$.
It will be useful computationally to consider the relation
\begin{equation}
(-1)^{\sigma_{k_i}(M)}\eta^{MN} = (-1)^{\sigma(M)}\eta^{MN}-2\frac{k_{i,A}^M k_{i,A}^N}{\Delta_i^2} \,,
\end{equation}
where we introduce
\begin{equation}
\sigma(M)=
\begin{cases}
1 &\text{if $M=0,\ldots,4$,}
\\
0 &\text{if $M=5,\ldots, 9$}\,,
\end{cases}
\end{equation}
and where we denote by $k_{i,\mathrm{A}}=(k_i^0,\dots,k_i^4,0,\dots,0)$ the
projection of $k_i$ onto the AdS$_5$ part, and $-\Delta_i^2=k_{i,\mathrm{A}}^2$.
We then have the following useful relations
\begin{align}
i\G^{0'}\G^{1'}\G^{2'}\G^{3'} &=\sfrac{1}{\Delta_i}\G^{0}\G^{1}\G^{2}\G^{3}\G^{4}\slk_{i,A}\,,
\\
\G^{4}\G^{3}\G^{2}\G^{1}\G^{0}\lrbrk{\G^{M_1}\dots\G^{M_n}}\G^{0}\G^{1}\G^{2}\G^{3}\G^{4}&=(-1)^{1+n+\sigma(M_1)+\dots+\sigma(M_n)}\G^{M_1}\dots\G^{M_n}\,.
\end{align}
Note that for the untwisted case, the same sixteen charges $Q_{\mathrm{L},\al \tilde a}-i\,Q_{\mathrm{R},\al \tilde a}$ and $Q_{\mathrm{L},\dot\al}^{\tilde a}+i\,Q_{\mathrm{R},\dot\al}^{\tilde a}$ annihilate both the vertex operator and its adjoint.

\section{Vertex function contractions}
\label{sec:Vertex}

\subsection{Fermion VEVs}

When computing correlators involving R-R vertices, we will need vacuum
expectation values for a number of fermion field combinations. Here we
precompute these vacuum expectation values, using bosonization
as outlined in \secref{sec:setup}.%
\footnote{We thank Oliver Schlotterer for mentioning that VEVs
involving terms $\psi^M(\slpsi\tw)^A$ can alternatively be computed
using the ``excited spin fields'' described in~\cite{Feng:2012bb}.}
In the following,
it is assumed that factors $\psi^M\bigbrk{\slpsi\tw}^A$
get contracted with massive polarization tensors
$t_{k,\tilde M\tilde A,MA}$ \eqref{eq:tmassive}, which obey
\begin{equation}
t_{k,\tilde M\tilde A,MA}\brk{\Gamma^M}^A_{\;\;B}=0\,,
\qquad
(\Gamma^{\tilde M}C)^{\tilde A\tilde B}t_{k,\tilde M\tilde B,MB}=0\,.
\label{eq:tgamma0}
\end{equation}
These relations hold for the twisted as well as for the untwisted tensors.
In computing the below correlators, we have added terms that vanish by
these relations in order to obtain a uniform dependence on the
worldsheet coordinates. It is also assumed that the term
$\psi^M\psi^N\psi^P$ is always contracted with the polarization tensor
\eqref{eq:alphaT}, such that all terms symmetric in $M,N,P$ are
discarded.
In the relations below, the ``$\simeq$'' signs mean that
the respective relations only hold under these assumptions. The
relevant VEVs are%
\footnote{The subscripts $1,2,3$ on the fields indicate
dependence on the worldsheet coordinate $z_{1,2,3}$.
We assume that all spin fields $\tw^A$ have positive chirality, and
we use the shorthand notation $\Gamma^{MNP\dots}$ for
$\Gamma^M\Gamma^N\Gamma^P\dots$.}
\begin{align}
&\Bigvev{
	\psi^M_{1}e^{-\phi_1}
	\tw^A_{2}e^{-\phi_2/2}
	\tw^B_{3}e^{-\phi_3/2}
}
=
\frac{1}{\sqrt{2}z_{12}z_{13}z_{23}}
\bigbrk{\Gamma^MC}^{AB},
\label{eq:corr133t1}
\\
&\Bigvev{
	\psi_1^M\psi_1^N\psi_1^Pe^{-\phi_1}
	\tw^A_{2}e^{-\phi_2/2}
	\tw^B_{3}e^{-\phi_3/2}
}
=
\frac{-1}{2\sqrt{2}z_{12}^2z_{13}^2}
\bigbrk{\Gamma^{MNP}C}^{AB},
\label{eq:corr233f1}
\\
&\Bigvev{
	\psi^M_{1}e^{-\phi_1}
	\psi_2^N\bigbrk{\slpsi_2\tw_2}^Ae^{-\phi_2/2}
	\tw^B_{3}e^{-\phi_3/2}
}
\simeq
\frac{2\sqrt{2}}{5z_{12}^2z_{23}^2}
\Bigbrk{
	\Gamma^{MN}C
	+8\eta^{MN}C
}^{AB},
\label{eq:corr133t2}
\\
&\Bigvev{
	\psi_1^M\psi_1^N\psi_1^Pe^{-\phi_1}
	\psi_2^Q\bigbrk{\slpsi_2\tw_2}^Ae^{-\phi_2/2}
	\tw^B_{3}e^{-\phi_3/2}
}
\simeq
\frac{-\sqrt{2}}{5z_{12}^3z_{13}z_{23}}
\Bigbrk{
	3\Gamma^{MNPQ}C
	+12\eta^{MQ}\Gamma^{NP}C
}^{AB},
\label{eq:corr233f2}
\\
&\Bigvev{
	\psi^M_{1}e^{-\phi_1}
	\psi_2^N\bigbrk{\slpsi_2\tw_2}^Ae^{-\phi_2/2}
	\psi_3^P\bigbrk{\slpsi_3\tw_3}^Be^{-\phi_3/2}
}
\simeq
\frac{16\sqrt{2}}{25z_{12}z_{13}z_{23}^3}
\cdot\nn\\&
\mspace{190mu}
\cdot
\Bigbrk{
	3\Gamma^{MNP}C
	-\eta^{MN}\Gamma^PC
	+5\eta^{MP}\Gamma^NC
	-25\eta^{NP}\Gamma^MC
}^{AB},
\label{eq:corr133t4}
\\
&\Bigvev{
	\psi_1^M\psi_1^N\psi_1^Pe^{-\phi_1}
	\psi_2^Q\bigbrk{\slpsi_2\tw_2}^Ae^{-\phi_3/2}
	\psi_3^R\bigbrk{\slpsi_3\tw_3}^Be^{-\phi_3/2}
}
\simeq
\frac{-64\sqrt{2}}{5z_{12}^2z_{13}^2z_{23}^2}
\cdot\nn\\&
\mspace{10mu}
\cdot
\Bigbrk{
	\sfrac{1}{40}\Gamma^{MNPQR}C
	-\sfrac{3}{8}\eta^{MR}\Gamma^{NPQ}C
	-\sfrac{21}{40}\eta^{MQ}\Gamma^{NPR}C
	-\sfrac{5}{8}\,\eta^{QR}\Gamma^{MNP}C
	-6\eta^{MQ}\eta^{NR}\Gamma^{P}C
}^{AB}.
\label{eq:corr233f4}
\end{align}
%

\subsection{Contractions with Three Chiral Primaries}
\label{sec:masslesscont}

In a three-point function of three massless vertices $W_{k_{1,2,3}}$,
we have $k_1^2=k_2^2=k_3^2=0$ and $k_i\cdot k_j=0$: All scalar
products among the momenta vanish.

\paragraph{\texorpdfstring{$\bigvev{W_1W_1W_1}$}{W1W1W1}.}

The correlator of three scalar level-zero NS-NS
vertices $W_1$ \eqref{eq:W1m1,eq:W10} reads%
\footnote{Here and everywhere below, we do not write the fermionic
ghosts $c$, $\tilde c$ explicitly, but we include their effect in the correlators.}
\begin{multline}
\bigvev{
W_{1,k_1}^{(-1,-1)}(z_1,\bar z_1)\,
W_{1,k_2}^{(-1,-1)}(z_2,\bar z_2)\,
W_{1,k_3}^{(0,0)}(z_3,\bar z_3)
}
=
-\gc^3\,\delta\bigbrk{{\textstyle\sum_jk_j}}
\cdot\\\cdot
\varepsilon_{k_1,M\tilde M}
\,\varepsilon_{k_2,N\tilde N}
\,\varepsilon_{k_3,P\tilde P}
\,\sapt
\bigbrk{\eta^{MN}k_1^P-k_1^N\eta^{MP}+k_2^M\eta^{NP}}
\bigbrk{\eta^{\tilde M\tilde N}k_1^{\tilde P}-k_1^{\tilde N}\eta^{\tilde M\tilde P}+k_2^{\tilde M}\eta^{\tilde N\tilde P}}\,.
\end{multline}
Using the twisted polarization tensors $\varepsilon_{\mathrm{T}}$, the
result vanishes because all scalar products $k_i\cdot k_j$ among the
three momenta vanish. But with untwisted polarization tensors, the
result becomes
\begin{multline}
\bigvev{
W_{1,k_1}(z_1,\bar z_1)\,
W_{1,k_2}(z_2,\bar z_2)\,
W_{1,k_3}(z_3,\bar z_3)
}
\\
=
\gc^3\,\delta\bigbrk{{\textstyle\sum_jk_j}}
\apt
\frac{-8(J_1^4+6J_1^2J_2^2+J_2^4+6J_1^2J_3^2+6J_2^2J_3^2+J_3^4)\tilde\alpha_1\tilde\alpha_2\tilde\alpha_3\tilde\Sigma}{J_1^2J_2^2J_3^2}\,,
\end{multline}
where $\tilde\alpha_j$, $\tilde\Sigma$ are defined in
\eqref{tildedefs}, and where $J_j=\sqrt{\brk{k_j-k_{j,\mathrm{A}}}^2}$
is the R-charge of the operator~$W_{1,k_j}$.%
\footnote{As before, we denote by
$k_{j,\mathrm{A}}=(k_j^0,\dots,k_j^4,0,\dots,0)$ the projection of
$k_j$ onto the AdS$_5$ part.}

\paragraph{\texorpdfstring{$\bigvev{W_1W_2W_2}$}{W1W2W2}.}

Using the VEV \eqref{eq:corr133t1},
the correlator of one flat-space scalar level-zero NS-NS vertex $W_1$
\eqref{eq:W1m1} with two
level-zero R-R vertices $W_2$ \eqref{eq:W2} reads
\begin{multline}
\bigvev{
W_{1,k_1}^{(-1,-1)}(z_1,\bar z_1)\,
W_{2,k_2}^{(-1/2,-1/2)}(z_2,\bar z_2)\,
W_{2,k_3}^{(-1/2,-1/2)}(z_3,\bar z_3)
}
\\
=
-\gc^3\,\delta\bigbrk{{\textstyle\sum_jk_j}}
\,\varepsilon_{k_1,M\tilde M}
\,t_{k_2,AB}
\,t_{k_3,CD}
\,\sfrac{1}{2}
\bigbrk{\Gamma^MC}^{BD}
\bigbrk{\Gamma^NC}^{AC}\,.
\end{multline}
Using the twisted polarization tensors $\varepsilon\indup{T}$,
$t\indup{T}$, the result is zero again, while with untwisted
polarizations it becomes%
\footnote{Because the theory is chiral, we use $\tr(\bfm{1})=16$.}
\begin{align}
\bigvev{
W_{1,k_1}(z_1,\bar z_1)\,
&W_{2,k_2}(z_2,\bar z_2)\,
W_{2,k_3}(z_3,\bar z_3)
}
\nn\\
&=
-
\gc^3\,\delta\bigbrk{{\textstyle\sum_jk_j}}
\,\varepsilon_{\mathrm{T},k_1,M\tilde M}
\apt
\frac{1}{\Delta_2\Delta_3}
\frac{1}{2}
\tr\bigsbrk{\Gamma^M\slk_3\slk_{3,\mathrm{A}}\Gamma^{43210}\Gamma^{\tilde M}\Gamma^{01234}\slk_{2,\mathrm{A}}\slk_2}
\nn\\
&=
-
\gc^3\,\delta\bigbrk{{\textstyle\sum_jk_j}}
\,\varepsilon_{\mathrm{T},k_1,M\tilde M}
\apt
\frac{1}{\Delta_2\Delta_3}
\frac{1}{2}
(-1)^{\sigma(\tilde M)}
\tr\bigsbrk{\Gamma^M\slk_3\slk_{3,\mathrm{A}}\Gamma^{\tilde M}\slk_{2,\mathrm{A}}\slk_2}
\nn\\
&=
\gc^3\,\delta\bigbrk{{\textstyle\sum_jk_j}}
\apt
\frac{-64(3J_1^2+J_2^2+J_3^2)\tilde\alpha_1\tilde\alpha_2\tilde\alpha_3\tilde\Sigma}{J_1^2J_2J_3}\,.
\end{align}
%

\paragraph{\texorpdfstring{$\bigvev{WWW}$}{WWW}.}

We are now in a position to assemble the correlator of three massless
chiral primaries. When all three operators are twisted the result vanishes trivially since all scalar products among the momenta vanish. As a check, we can also compute the correlator with two twisted operators and again one obtains that it vanishes, which is due to a supersymmetric Ward identity. Finally the correlator for three primaries \eqref{masslessnormvert}  yields
\eqref{eq:www}
\begin{equation}
\bigvev{W_{k_1}(z_1,\bar z_1)W_{k_2}(z_2,\bar z_2)W_{k_3}(z_3,\bar z_3)}
=
\gc^3\,\delta\bigbrk{{\textstyle\sum_jk_j}}
\apt
\frac{2\tilde\alpha_1\tilde\alpha_2\tilde\alpha_3\tilde\Sigma^5}{J_1^2J_2^2J_3^2}\,.
\label{eq:WWW}
\end{equation}
%

\subsection{Contractions with Two Chiral Primaries}
\label{sec:contract2chiral}

With two massless vertices, $W_{k_{1,2}}$ and the level-one vertex $V_{k_3}$ we have
$k_1^2=k_2^2=0$, $k_3^2=-4/\ap$, $k_{1,2}\cdot k_3=2/\ap$ and $k_1\cdot k_2=-2/\ap$.

\paragraph{\texorpdfstring{$\bigvev{W_1W_1V_1}$}{W1W1V1}.}

Both the left and right moving part of the correlator
$\bigvev{
W_{1,k_1}^{(-1,-1)}\,
W_{1,k_2}^{(-1,-1)}\,
V_{1,k_3}^{(0,0)}
}$
are split into three parts corresponding to the terms in
$V_{1,k_3}^{(0,0)}$.
Using the symmetries of the polarization tensors, combining the three
parts and multiplying contractions of the left-moving with the
right-moving part, we obtain%
\footnote{We do not denote
the fermionic ghosts $c$, $\tilde c$ explicitly, but include their effect in the
result.}
\begin{multline}
\bigvev{
W_{1,k_1}^{(-1,-1)}(z_1,\bar z_1)\,
W_{1,k_2}^{(-1,-1)}(z_2,\bar z_2)\,
V_{1,k_3}^{(0,0)}(z_3,\bar z_3)
}
\\
=
-\gc^3\,\delta\bigbrk{{\textstyle\sum_jk_j}}
\,\eps^{k_1}_{M,\tilde M}
\,\eps^{k_2}_{N,\tilde N}
\,\eps^{k_3}_{RS,\tilde R\tilde S}
\,X^{MNRS}
X^{\tilde M\tilde N\tilde R\tilde S}\,,
\end{multline}
where
\begin{equation}
X^{MNRS}
=
\apt
\Bigbrk{
\eta^{MN} k_1^R k_1^S+\stap\eta^{MS}\eta^{NR}+\bigbrk{\eta^{MR}k_3^N-\eta^{NR}k_3^M} k_1^S
}\,.
\end{equation}
Expanding the twisted polarization tensors and performing the index
contractions gives
\begin{equation}
\bigvev{
W_{1,\mathrm{T},k_1}^{(-1,-1)}(z_1,\bar z_1)\,
W_{1,\mathrm{T},k_2}^{(-1,-1)}(z_2,\bar z_2)\,
V_{1,\mathrm{T},k_3}^{(0,0)}(z_3,\bar z_3)
}
\\
=
-2^2\times 3^2\,\gc^3\,\delta\bigbrk{{\textstyle\sum_jk_j}}\,.
\label{eq:W1W1V1twisted}
\end{equation}
When contracting with the untwisted polarization tensors, the result
becomes a lengthy rational function of $\Delta_{1,2,3}$ that we do not quote here.

\paragraph{\texorpdfstring{$\bigvev{W_1W_1V_2}$}{W1W1V2}.}

For the correlator
$\bigvev{
W_{1,k_1}^{(-1,-1)}\,
W_{1,k_2}^{(-1,-1)}\,
V_{2,k_3}^{(0,0)}
}$,
we have to consider only the first term of $V_{2,k_3}^{(0,0)}$ since for the second there are $\psi$'s which cannot be contracted.
Assuming antisymmetry of the polarization tensor $\alpha$, the
result is
\begin{multline}
\bigvev{
W_{1,k_1}^{(-1,-1)}(z_1,\bar z_1)\,
W_{1,k_2}^{(-1,-1)}(z_2,\bar z_2)\,
V_{2,k_3}^{(0,0)}(z_3,\bar z_3)
}
\\
=
-\gc^3\,\delta\bigbrk{{\textstyle\sum_jk_j}}
\,\eps^{k_1}_{M,\tilde M }
\,\eps^{k_2}_{N,\tilde N}
\,\alpha^{k_3}_{RST,\tilde R\tilde S\tilde T}
\,\sapt
36
\eta^{MT}\eta^{NS}k_1^R
\eta^{\tilde M\tilde T}\eta^{\tilde N\tilde S}k_1^{\tilde R}\,.
\end{multline}
Expanding the twisted polarization tensors and performing the index contractions gives
\begin{equation}
\bigvev{
W_{1,\mathrm{T},k_1}^{(-1,-1)}(z_1,\bar z_1)\,
W_{1,\mathrm{T},k_2}^{(-1,-1)}(z_2,\bar z_2)\,
V_{2,\mathrm{T},k_3}^{(0,0)}(z_3,\bar z_3)
}
\\
=
-2^2\times7
\,\gc^3\,\delta\bigbrk{{\textstyle\sum_jk_j}}\,.
\label{eq:W1W1V2twisted}
\end{equation}
When contracting with the untwisted polarization tensors, the result
again becomes a complicated rational function of $\Delta_{1,2,3}$ that we do not quote here.

\paragraph{\texorpdfstring{$\bigvev{V_1W_2W_2}$}{V1W2W2}.}

Using the VEV \eqref{eq:corr133t1},
we can compute the correlator
\begin{multline}
\bigvev{
V_{1,k_1}^{(-1,-1)}(z_1,\bar z_1)\,
W_{2,k_2}^{(-1/2,-1/2)}(z_2,\bar z_2)\,
W_{2,k_3}^{(-1/2,-1/2)}(z_3,\bar z_3)
}
\\
=
-\gc^3\,\delta\bigbrk{{\textstyle\sum_jk_j}}
\,\eps^{k_1}_{MN,\tilde M\tilde N}
\,\frac{1}{2}
\Tr\Bigsbrk{
t^{k_2}
X^{MN}
\bigbrk{t^{k_3}}\transpose
\bigbrk{X^{\tilde M\tilde N}}\transpose
}\,,
\end{multline}
where superscript $\transpose$ denotes transposition in spinor indices, and
\begin{equation}
X^{MN}
=
\sqrt\apt
\,k_2^N
\Gamma^MC
\end{equation}
are the left/right-moving field contractions.
Expanding the twisted polarization tensors and performing the index contractions and matrix algebra gives
\begin{equation}
\bigvev{
V_{1,\mathrm{T},k_1}^{(-1,-1)}(z_1,\bar z_1)\,
W_{2,\mathrm{T},k_2}^{(-1/2,-1/2)}(z_2,\bar z_2)\,
W_{2,\mathrm{T},k_3}^{(-1/2,-1/2)}(z_3,\bar z_3)
}
\\
=
-2^4\,\gc^3\,\delta\bigbrk{{\textstyle\sum_jk_j}}\,.
\label{eq:V1W2W2twisted}
\end{equation}
Contracting with the untwisted polarization tensors gives an expression that we will refrain from quoting here.

\paragraph{\texorpdfstring{$\bigvev{V_2W_2W_2}$}{V2W2W2}.}

Using the VEV \eqref{eq:corr233f1},
the correlator is given by
\begin{multline}
\bigvev{
V_{2,k_1}^{(-1,-1)}(z_1,\bar z_1)\,
W_{2,k_2}^{(-1/2,-1/2)}(z_2,\bar z_2)\,
W_{2,k_3}^{(-1/2,-1/2)}(z_3,\bar z_3)
}
\\
=
-\gc^3\,\delta\bigbrk{{\textstyle\sum_jk_j}}
\alpha^{k_1}_{MNP,\tilde M\tilde N\tilde P}
\,\frac{1}{2}
\Tr\Bigsbrk{
t^{k_2}
X^{MNP}
\bigbrk{t^{k_3}}\transpose
\bigbrk{X^{\tilde M\tilde N\tilde P}}\transpose
}\,,
\end{multline}
where
\begin{equation}
X^{MNP}
=
\half
\Gamma^M\Gamma^N\Gamma^PC
\end{equation}
are the left/right-moving field contractions.
Expanding the twisted polarization tensors and performing the index contractions and matrix algebra gives
\begin{equation}
\bigvev{
V_{2,\mathrm{T},k_1}^{(-1,-1)}(z_1,\bar z_1)\,
W_{2,\mathrm{T},k_2}^{(-1/2,-1/2)}(z_2,\bar z_2)\,
W_{2,\mathrm{T},k_3}^{(-1/2,-1/2)}(z_3,\bar z_3)
}
\\
=
-2^4\times 7
\,\gc^3\,\delta\bigbrk{{\textstyle\sum_jk_j}}\,.
\label{eq:V2W2W2twisted}
\end{equation}
Once again, contracting with the untwisted polarization tensors gives a complicated expression that we will not write here.

\paragraph{\texorpdfstring{$\bigvev{W_1W_2V_3}$}{W1W2V3}.}

Using the VEVs \eqref{eq:corr133t1,eq:corr133t2},
we compute the last correlator
\begin{multline}
\bigvev{
W_{1,k_1}^{(-1,-1)}(z_1,\bar z_1)\,
W_{2,k_2}^{(-1/2,-1/2)}(z_2,\bar z_2)\,
V_{3,k_3}^{(-1/2,-1/2)}(z_3,\bar z_3)
}
\\
=
-\gc^3\,\delta\bigbrk{{\textstyle\sum_jk_j}}
\eps^{k_1}_{M,\tilde M}
\,\frac{1}{2}
\Tr\Bigsbrk{
t^{k_2}
X^{MN}
\bigbrk{t^{k_3}_{\tilde N,N}}\transpose
\bigbrk{X^{\tilde M\tilde N}}\transpose
}\,,
\end{multline}
where
\begin{equation}
X^{MN}
=
\sqrt\apt
\bigbrk{
k_1^N\Gamma^MC
+\eta^{MN}\slk_3C
}
\end{equation}
are the left/right-moving field contractions, which have been
simplified using \eqref{eq:tgamma0} as well as the fact that all contractions of $k_j$ with $\eps_{k_j}$, $t_{k_j}$ vanish.
Expanding the twisted polarization tensors and performing the index contractions and matrix algebra gives
\begin{equation}
\bigvev{
W_{1,\mathrm{T},k_1}^{(-1,-1)}(z_1,\bar z_1)\,
W_{2,\mathrm{T},k_2}^{(-1/2,-1/2)}(z_2,\bar z_2)\,
V_{3,\mathrm{T},k_3}^{(-1/2,-1/2)}(z_3,\bar z_3)
}
\\
=
2^7\,\gc^3\,\delta\bigbrk{{\textstyle\sum_jk_j}}\,.
\label{eq:W1W2V3twisted}
\end{equation}
Contracting with the untwisted polarization tensors gives a lengthy expression that we will refrain from writing here.

\paragraph{\texorpdfstring{$\bigvev{WWV}$}{WWV}.}

We can now assemble the different parts of the correlator with two massless vertices $W$ \eqref{masslessnormvert}  and one massive vertex $V$ \eqref{massivenormvert}. As a check, we first compute the correlators with two and three twisted vertex operators, and we obtain that all of them vanish, which is explained by a supersymmetric Ward identity.
For three untwisted vertices, meaning they are all primaries, all the complicated results of the different parts combine into the simple expression
\eqref{eq:wwv}
\be
\vev{W_{k_1}(z_1,\bar z_1)W_{k_2}(z_2,\bar z_2)V_{k_3}(z_3,\bar z_3)}
=
\gc^3\,\delta\bigbrk{{\textstyle\sum_jk_j}}
\,\ap^2\,\frac{\al_1^2\al_2^2\Sigma^4\tilde\al_3^2\tilde\Sigma^2}{\Delta_1^2\Delta_2^2\Delta_3^4}\,.
\ee

\subsection{Contractions with One Chiral Primary}
\label{sec:contract1chiral}

With two level-one vertices, $V_{k_{1,2}}$, and the massless vertex $W_{k_3}$ we have
$k_1^2=k_2^2=-4/\ap$, $k_3^2=0$, $k_{1,2}\cdot k_3=0$ and $k_1\cdot k_2=4/\ap$.

\paragraph{\texorpdfstring{$\bigvev{V_1V_1W_1}$}{V1V1W1}.}

The left and right-moving parts of the correlator
$\bigvev{
V_{1,k_1}^{(-1,-1)}\,
V_{1,k_2}^{(-1,-1)}\,
W_{1,k_3}^{(0,0)}
}$
are split into two parts corresponding to the terms in
$W_{1,k_3}^{(0,0)}$.
Using the symmetries of the polarization tensors, combining those parts and multiplying left and right-moving contractions we obtain
\footnote{Once again, we do not denote
the fermionic ghosts $c$, $\tilde c$ explicitly, but include their effect in the
result.}
\begin{multline}
\bigvev{
V_{1,k_1}^{(-1,-1)}(z_1,\bar z_1)\,
V_{1,k_2}^{(-1,-1)}(z_2,\bar z_2)\,
W_{1,k_3}^{(0,0)}(z_3,\bar z_3)
}
\\
=
-\gc^3\,\delta\bigbrk{{\textstyle\sum_jk_j}}
\,\eps^{k_1}_{MN,\tilde M\tilde N}
\,\eps^{k_2}_{PQ,\tilde P\tilde Q}
\,\eps^{k_3}_{R,\tilde R}
\,X^{MNPQR}
X^{\tilde M\tilde N\tilde P\tilde Q\tilde R}\,,
\end{multline}
where
\begin{multline}
X^{MNPQR}
=
\sqrt{\sapt}
\eta^{MP}
\bigbrk{
\eta^{NQ}k_1^R
+2\eta^{QR}k_2^N
+2\eta^{NR}k_3^Q
+\sapt k_2^N k_3^Q k_1^R
}
\\
+
\bigbrk{\sapt}^{3/2}
\bigbrk{\eta^{MR}k_3^P+\eta^{PR}k_2^M}k_2^Nk_3^Q
\,.
\end{multline}
Expanding the twisted polarization tensors and performing the index
contractions gives
\begin{equation}
\bigvev{
V_{1,\mathrm{T},k_1}^{(-1,-1)}(z_1,\bar z_1)\,
V_{1,\mathrm{T},k_2}^{(-1,-1)}(z_2,\bar z_2)\,
W_{1,\mathrm{T},k_3}^{(0,0)}(z_3,\bar z_3)
}
\\
=
2^3\times11
\,\gc^3\,\delta\bigbrk{{\textstyle\sum_jk_j}}\,.
\label{eq:V1V1W1twisted}
\end{equation}
When contracting with the untwisted polarization tensors, the result
becomes a lengthy rational function of $\Delta_{1,2,3}$ that we do not quote here.

\paragraph{\texorpdfstring{$\bigvev{V_2V_2W_1}$}{V2V2W1}.}

As before, there are two parts in the correlator
$\bigvev{
V_{2,k_1}^{(-1,-1)}\,
V_{2,k_2}^{(-1,-1)}\,
W_{1,k_3}^{(0,0)}
}$.
Assuming antisymmetry of the polarization tensor $\alpha$, they combine into
\begin{multline}
\bigvev{
V_{2,k_1}^{(-1,-1)}(z_1,\bar z_1)\,
V_{2,k_2}^{(-1,-1)}(z_2,\bar z_2)\,
W_{1,k_3}^{(0,0)}(z_3,\bar z_3)
}
=
-\gc^3\,\delta\bigbrk{{\textstyle\sum_jk_j}}
\,\alpha^{k_1}_{MNP,\tilde M\tilde N\tilde N}
\,\alpha^{k_2}_{QRS,\tilde Q\tilde R\tilde S}
\,\eps^{k_3}_{T,\tilde T}
\cdot\\\cdot
\apt
36
\eta^{NS}\eta^{PR}
\Bigsbrk{
k_1^T\eta^{MQ}
+
3\Bigbrk{\eta^{MT}k_3^Q-\eta^{QT}k_3^M}
}
\eta^{\tilde N\tilde S}\eta^{\tilde P\tilde R}
\Bigsbrk{
k_1^{\tilde T}\eta^{\tilde M\tilde Q}
+
3\Bigbrk{\eta^{\tilde M\tilde T}k_3^{\tilde Q}-\eta^{\tilde Q\tilde T}k_3^{\tilde M}}
}
\,.
\end{multline}
Expanding the twisted polarization tensors and performing the index
contractions gives
\begin{equation}
\bigvev{
V_{2,\mathrm{T},k_1}^{(-1,-1)}(z_1,\bar z_1)\,
V_{2,\mathrm{T},k_2}^{(-1,-1)}(z_2,\bar z_2)\,
W_{1,\mathrm{T},k_3}^{(0,0)}(z_3,\bar z_3)
}
\\
=
2^3\times3\times7
\,\gc^3\,\delta\bigbrk{{\textstyle\sum_jk_j}}\,.
\label{eq:V2V2W1twisted}
\end{equation}
Contracting with the untwisted polarization tensors, the result
becomes a complicated function of $\Delta_{1,2,3}$ that we will not quote here.

\paragraph{\texorpdfstring{$\bigvev{V_2V_1W_1}$}{V2V1W1}.}

In the case of
$\bigvev{
V_{2,k_1}^{(-1,-1)}\,
V_{1,k_2}^{(-1,-1)}\,
W_{1,k_3}^{(0,0)}
}$,
we have to consider only the second term of $W_{1,k_3}^{(0,0)}$ since for the first there are $\psi$'s which cannot be contracted. Then, assuming antisymmetry of the polarization tensor $\alpha$, the correlator is given by
\begin{multline}
\bigvev{
V_{2,k_1}^{(-1,-1)}(z_1,\bar z_1)\,
V_{1,k_2}^{(-1,-1)}(z_2,\bar z_2)\,
W_{1,k_3}^{(0,0)}(z_3,\bar z_3)
}
\\
=
-\gc^3\,\delta\bigbrk{{\textstyle\sum_jk_j}}
\,\alpha^{k_1}_{MNP,\tilde M\tilde N\tilde N}
\,\eps^{k_2}_{RS,\tilde R\tilde S}
\,\eps^{k_3}_{T,\tilde T}
\,\lrbrk{\sapt}^2
36k_3^S k_3^N\eta^{MT}\eta^{PR} k_3^{\tilde S} k_3^{\tilde N}\eta^{\tilde M\tilde T}\eta^{\tilde P\tilde R}
\,.
\end{multline}
Expanding the twisted polarization tensors and performing the index
contractions gives
\begin{equation}
\bigvev{
V_{2,\mathrm{T},k_1}^{(-1,-1)}(z_1,\bar z_1)\,
V_{1,\mathrm{T},k_2}^{(-1,-1)}(z_2,\bar z_2)\,
W_{1,\mathrm{T},k_3}^{(0,0)}(z_3,\bar z_3)
}
=
0\,.
\label{eq:V2V1W1twisted}
\end{equation}
As before, we do not quote the result of contracting with the untwisted polarization tensors.

\paragraph{\texorpdfstring{$\bigvev{V_1W_2V_3}$}{V1W2V3}.}

Using the VEVs \eqref{eq:corr133t1,eq:corr133t2},
we can compute the correlator
\begin{multline}
\bigvev{
V_{1,k_1}^{(-1,-1)}(z_1,\bar z_1)\,
W_{2,k_2}^{(-1/2,-1/2)}(z_2,\bar z_2)\,
V_{3,k_3}^{(-1/2,-1/2)}(z_3,\bar z_3)
}
\\
=
-\gc^3\,\delta\bigbrk{{\textstyle\sum_jk_j}}
\,\eps^{k_1}_{MN,\tilde M\tilde N}
\frac{1}{2}\Tr\Bigsbrk{
t^{k_2}
X^{MNP}
\bigbrk{t^{k_3}_{\tilde P,P}}\transpose
\bigbrk{X^{\tilde M\tilde N\tilde P}}\transpose
}\,,
\end{multline}
where
\begin{equation}
X^{MNP}
=
\bigbrk{
\eta^{NP}+\sapt k_2^N k_1^P
}
\Gamma^MC
+
\sapt k_2^N\eta^{MP}\slk_3C
\end{equation}
are the left/right-moving field contractions, which have been
simplified using \eqref{eq:tgamma0} as well as the symmetry of
$\varepsilon_{k_1}$ and the fact that all contractions of $k_j$ with
$\varepsilon_{k_j}$, $t_{k_j}$ vanish.
Expanding the twisted polarization tensors and performing the index contractions and matrix algebra gives
\begin{equation}
\bigvev{
V_{1,\mathrm{T},k_1}^{(-1,-1)}(z_1,\bar z_1)\,
W_{2,\mathrm{T},k_2}^{(-1/2,-1/2)}(z_2,\bar z_2)\,
V_{3,\mathrm{T},k_3}^{(-1/2,-1/2)}(z_3,\bar z_3)
}
=
0\,.
\label{eq:V1W2V3twisted}
\end{equation}
Once again, contracting with the untwisted polarization tensors gives an expression that we will refrain from quoting here.

\paragraph{\texorpdfstring{$\bigvev{V_2W_2V_3}$}{V2W2V3}.}

Using the VEVs \eqref{eq:corr233f1,eq:corr233f2},
the correlator is given by
\begin{multline}
\bigvev{
V_{2,k_1}^{(-1,-1)}(z_1,\bar z_1)\,
W_{2,k_2}^{(-1/2,-1/2)}(z_2,\bar z_2)\,
V_{3,k_3}^{(-1/2,-1/2)}(z_3,\bar z_3)
}
\\
=
-\gc^3\,\delta\bigbrk{{\textstyle\sum_jk_j}}
\,\alpha^{k_1}_{MNP,\tilde M\tilde N\tilde P}
\,\frac{1}{2}
\Tr\Bigsbrk{
t^{k_2}
X^{MNPR}
\bigbrk{t^{k_3}_{\tilde R,R}}\transpose
\bigbrk{X^{\tilde M\tilde N\tilde P\tilde R}}\transpose
}\,,
\end{multline}
where
\begin{equation}
X^{MNPR}
=
\sqrt{\apt}
\frac{1}{2}
k_1^R
\Gamma^M\Gamma^N\Gamma^PC
+
\sqrt{\apt}\frac{3}{2}\eta^{MR}
\Gamma^N\Gamma^P\slk_3C
\end{equation}
are the left/right-moving field contractions, which have been
simplified using \eqref{eq:tgamma0} as well as the symmetry of
$\alpha_{k_1}$ and the fact that all contractions of $k_j$ with
$\alpha_{k_j}$, $t_{k_j}$ vanish.
Expanding the twisted polarization tensors and performing the index contractions and matrix algebra gives
\begin{equation}
\bigvev{
V_{2,\mathrm{T},k_1}^{(-1,-1)}(z_1,\bar z_1)\,
W_{2,\mathrm{T},k_2}^{(-1/2,-1/2)}(z_2,\bar z_2)\,
V_{3,\mathrm{T},k_3}^{(-1/2,-1/2)}(z_3,\bar z_3)
}
=
0\,.
\label{eq:V2W2V3twisted}
\end{equation}
Contracting with the untwisted polarization tensors gives a complicated function of $\Delta_{1,2,3}$ that we will refrain from quoting here.

\paragraph{\texorpdfstring{$\bigvev{W_1V_3V_3}$}{W1V3V3}.}

Using the VEVs \eqref{eq:corr133t1,eq:corr133t2,eq:corr133t4},
the last correlator is
\begin{multline}
\bigvev{
W_{1,k_1}^{(-1,-1)}(z_1,\bar z_1)\,
V_{3,k_2}^{(-1/2,-1/2)}(z_2,\bar z_2)\,
V_{3,k_3}^{(-1/2,-1/2)}(z_3,\bar z_3)
}
\\
=
-\gc^3\,\delta\bigbrk{{\textstyle\sum_jk_j}}
\,\eps^{k_1}_{M,\tilde M}
\,\frac{1}{2}\Tr\Bigsbrk{
t^{k_2}_{\tilde N,N}
X^{MNP}
\bigbrk{t^{k_3}_{\tilde P,P}}\transpose
\bigbrk{X^{\tilde M\tilde N\tilde P}}\transpose
}\,,
\end{multline}
where
\begin{equation}
X^{MNP}
=
\bigbrk{
\eta^{NP}+\sapt k_3^N k_1^P
}
\Gamma^MC
+
\sapt k_3^N\eta^{MP}
\slk_3C
-
\sapt k_1^P\eta^{MN}
\slk_2C
+
\sapt\sfrac{1}{2}\eta^{NP}
\slk_2\Gamma^M\slk_3C
\end{equation}
are the left/right-moving field contractions, which have been
simplified using \eqref{eq:tgamma0} as well as the fact that all contractions of $k_j$ with
$\eps_{k_j}$, $t_{k_j}$ vanish.
Expanding the twisted polarization tensors and performing the index contractions and matrix algebra gives
\begin{equation}
\bigvev{
W_{1,\mathrm{T},k_1}^{(-1,-1)}(z_1,\bar z_1)\,
V_{3,\mathrm{T},k_2}^{(-1/2,-1/2)}(z_2,\bar z_2)\,
V_{3,\mathrm{T},k_3}^{(-1/2,-1/2)}(z_3,\bar z_3)
}
=
-2^9\,\gc^3\,\delta\bigbrk{{\textstyle\sum_jk_j}}\,.
\label{eq:W1V3V3twisted}
\end{equation}
Again, we do not write here the result of contracting with the untwisted polarization tensors.

\paragraph{\texorpdfstring{$\bigvev{VVW}$}{VVW}.}

We now can assemble all the different parts of the correlator of a
massless vertex $W$ \eqref{masslessnormvert} and two massive vertices
$V$  \eqref{massivenormvert}. When two or three of them are twisted,
we find once again that the results vanish, which matches the supersymmetric Ward identity.
For three primaries, the correlator becomes \eqref{eq:vvw}
\begin{multline}
\vev{V_{k_1}(z_1,\bar z_1)V_{k_2}(z_2,\bar z_2)W_{k_3}(z_3,\bar z_3)}=
\gc^3\,\delta\bigbrk{{\textstyle\sum_jk_j}}
\frac{\Sigma^4}{\Delta_1^4\Delta_2^4\Delta_3^2}
\cdot\\\cdot
\Bigbrk{
-\half\Omega_1^2\al_3^4+\al_3^2\lrbrk{3\Omega_1^2+2\Omega_1\al_3+\al_3^2}\Omega_4
-\half\Omega_1(\Omega_1-4\al_3)\Omega_4^2+\Omega_4^3
}\,.
\end{multline}
where
\begin{equation}
\Omega_1=\alpha_1+\alpha_2\,,
\qquad
\Omega_4=\ap\alpha_1\alpha_2\alpha_3\Sigma\,.
\end{equation}
%

\subsection{Contractions with Zero Chiral Primaries}
\label{sec:contract0chirals}

With three level-one vertices, we have
$k_j^2=-4/\ap$, and $k_i\cdot k_j=2/\ap$ for $i\neq j$.

\paragraph{\texorpdfstring{$\bigvev{V_1V_1V_1}$}{V1V1V1}.}

The correlator of three vertices $V_1$ \eqref{eq:V1m1,eq:V100}
evaluates to%
\footnote{As before, we do not denote the fermionic ghosts $c$,
$\tilde c$ explicitly, but include their effect in the result.}
\begin{multline}
\bigvev{
V_{1,k_1}^{(-1,-1)}(z_1,\bar z_1)\,
V_{1,k_2}^{(-1,-1)}(z_2,\bar z_2)\,
V_{1,k_3}^{(0,0)}(z_3,\bar z_3)
}
\\
=
-\gc^3\,\delta\bigbrk{{\textstyle\sum_jk_j}}
\varepsilon^{k_1}_{MN,\tilde M\tilde N}
\,\varepsilon^{k_2}_{PQ,\tilde P\tilde Q}
\,\varepsilon^{k_3}_{RS,\tilde R\tilde S}
\,X^{MNPQRS}X^{\tilde M\tilde N\tilde P\tilde Q\tilde R\tilde S}\,,
\end{multline}
where, using the antisymmetry of the polarization tensor
$\varepsilon_{MN,\tilde M\tilde N}$ as well as
$k_j\cdot\varepsilon^{k_j}=0$,
\begin{multline}
X^{MNPQRS}
=
\half\eta^{MP}
\Bigbrk{
\eta^{NR}\bigbrk{
\eta^{QS}
-\sapt3k_1^Qk_1^S
}
-
\sapt\bigbrk{
\eta^{NQ}
-\sapt k_3^Nk_3^Q
}k_1^Rk_2^S
}
\\
+\bigbrk{5\text{ permutations of }\bigbrc{\brk{M,N,k_1},\brk{P,Q,k_2},\brk{R,S,k_3}}}
\,.
\end{multline}
Contracting with the twisted polarization tensors
$\varepsilon_{\mathrm{T}}$ \eqref{eq:epsmassive} gives
\begin{equation}
\bigvev{
V_{1,\mathrm{T},k_1}^{(-1,-1)}(z_1,\bar z_1)\,
V_{1,\mathrm{T},k_2}^{(-1,-1)}(z_2,\bar z_2)\,
V_{1,\mathrm{T},k_3}^{(0,0)}(z_3,\bar z_3)
}
=
-2^4\times3\times53
\,\gc^3\,\delta\bigbrk{{\textstyle\sum_jk_j}}\,.
\label{eq:V1TV1TV1T}
\end{equation}
When contracting with the untwisted polarization tensors, the result
becomes a lengthy rational function of $\Delta_1$, $\Delta_2$,
$\Delta_3$ (or equivalently $J_1$, $J_2$, $J_3$) that we do not quote
here.
For $J_i\ll\sqrt{1/\ap}$ (i.e.\
$\Delta_i\approx\sqrt{4/\ap}$), it becomes exactly
\eqref{eq:V1TV1TV1T}, plus $\order{J_i^2\ap}$ terms. This can be
understood by noting that all terms in the contraction contain an even
number of untwisted index pairs.

\paragraph{\texorpdfstring{$\bigvev{V_1V_1V_2}$}{V1V1V2}.}

In the correlator
$\bigvev{
V_{1,k_1}^{(-1,-1)}\,
V_{1,k_2}^{(-1,-1)}\,
V_{2,k_3}^{(0,0)}
}$,
only the first term in $V_{2,k_3}^{(0,0)}$ \eqref{eq:V200} contributes, since the second
term yields an odd number of $\psi$'s, which cannot be fully contracted.
Assuming antisymmetry of the polarization tensor $\alpha$, the
result is
\begin{multline}
\bigvev{
V_{1,k_1}^{(-1,-1)}(z_1,\bar z_1)\,
V_{1,k_2}^{(-1,-1)}(z_2,\bar z_2)\,
V_{2,k_3}^{(0,0)}(z_3,\bar z_3)
}
\\
=
-\gc^3\,\delta\bigbrk{{\textstyle\sum_jk_j}}
\varepsilon^{k_1}_{MN,\tilde M\tilde N}
\,\varepsilon^{k_2}_{PQ,\tilde P\tilde Q}
\,\alpha^{k_3}_{RST,\tilde R\tilde S\tilde T}
\,X^{MNPQRST}
X^{\tilde M\tilde N\tilde P\tilde Q\tilde R\tilde S\tilde T}\,,
\end{multline}
where
\begin{equation}
X^{MNPQRST}
=
\sqrt{\sapt}
\,6\eta^{MR}\eta^{PS}
\bigbrk{\eta^{NT}k_1^Q-\eta^{NQ}k_1^T-\eta^{QT}k_2^N+\sapt k_2^Nk_1^Qk_1^T} \,.
\end{equation}
Contracting with the twisted polarization tensors
$\varepsilon\indup{T}$ \eqref{eq:epsmassive} and $\alpha\indup{T}$
\eqref{eq:alphaT} gives
\begin{equation}
\bigvev{
V_{1,\mathrm{T},k_1}^{(-1,-1)}(z_1,\bar z_1)\,
V_{1,\mathrm{T},k_2}^{(-1,-1)}(z_2,\bar z_2)\,
V_{2,\mathrm{T},k_3}^{(0,0)}(z_3,\bar z_3)
}
\\
=
2\times3\times7^2
\,\gc^3\,\delta\bigbrk{{\textstyle\sum_jk_j}}\,.
\label{eq:V1V1V2twisted}
\end{equation}
The result for untwisted vertices with general $J_i$ is again lengthy
and we refrain from quoting it.
For $J_i\ll\sqrt{1/\ap}$, it becomes
exactly the negative of \eqref{eq:V1V1V2twisted}, plus $\order{J_i^2\ap}$ terms, which follows
from the fact that all terms in the contraction contain an odd number
of untwisted index pairs.

\paragraph{\texorpdfstring{$\bigvev{V_2V_2V_1}$}{V2V2V1}.}

Performing the field contractions, the correlator
$\bigvev{
V_{2,k_1}
V_{2,k_2}
V_{1,k_3}
}$
becomes
\begin{multline}
\bigvev{
V_{2,k_1}^{(-1,-1)}(z_1,\bar z_1)\,
V_{2,k_2}^{(-1,-1)}(z_2,\bar z_2)\,
V_{1,k_3}^{(0,0)}(z_3,\bar z_3)
}
\\
=
-\gc^3\,\delta\bigbrk{{\textstyle\sum_jk_j}}
\alpha^{k_1}_{MNP,\tilde M\tilde N\tilde P}
\,\alpha^{k_2}_{QRS,\tilde Q\tilde R\tilde S}
\,\varepsilon^{k_3}_{TU,\tilde T\tilde U}
\,X^{MNPQRSTU}
X^{\tilde M\tilde N\tilde P\tilde Q\tilde R\tilde S\tilde T\tilde U}\,,
\end{multline}
where
\begin{equation}
X^{MNPQRSTU}
=
6\eta^{NR}\eta^{PS}
\Bigbrk{
3\eta^{MU}\bigbrk{\eta^{QT}-\sapt k_1^Qk_1^T}
+\sapt\bigbrk{\eta^{MQ}k_1^U+3\eta^{QU}k_2^M}k_1^T
}
\,.
\end{equation}
Expanding the twisted polarization tensors and performing the index
contractions gives
\begin{equation}
\bigvev{
V_{2,\mathrm{T},k_1}^{(-1,-1)}(z_1,\bar z_1)\,
V_{2,\mathrm{T},k_2}^{(-1,-1)}(z_2,\bar z_2)\,
V_{1,\mathrm{T},k_3}^{(0,0)}(z_3,\bar z_3)
}
=
-2^2\times3\times7\times19
\,\gc^3\,\delta\bigbrk{{\textstyle\sum_jk_j}}\,.
\label{eq:V2V2V1twisted}
\end{equation}
We refrain from writing out the lengthy result for untwisted vertices.
Again, for $J_i\ll\sqrt{1/\ap}$, it becomes exactly
\eqref{eq:V2V2V1twisted}, plus $\order{J_i^2\ap}$ terms, because each term contains an even number of
untwisted index contractions.

\paragraph{\texorpdfstring{$\bigvev{V_2V_2V_2}$}{V2V2V2}.}

Performing the field contractions, the correlator
$\bigvev{
V_{2,k_1}\,
V_{2,k_2}\,
V_{2,k_3}
}$
becomes (using the antisymmetry of the polarization tensors $\alpha$)
\begin{multline}
\bigvev{
V_{2,k_1}^{(-1,-1)}(z_1,\bar z_1)\,
V_{2,k_2}^{(-1,-1)}(z_2,\bar z_2)\,
V_{2,k_3}^{(0,0)}(z_3,\bar z_3)
}
\\
=
-\gc^3\,\delta\bigbrk{{\textstyle\sum_jk_j}}
\alpha^{k_1}_{MNP,\tilde M\tilde N\tilde P}
\,\alpha^{k_2}_{QRS,\tilde Q\tilde R\tilde S}
\,\alpha^{k_3}_{TUV,\tilde T\tilde U\tilde V}
\,X^{MNPQRSTUV}
X^{\tilde M\tilde N\tilde P\tilde Q\tilde R\tilde S\tilde T\tilde U\tilde V}\,,
\end{multline}
where
\begin{equation}
X^{MNPQRSTUV}
=
\,2^2\,3^3
\eta^{MQ}\eta^{SV}
\sqrt{\sapt}
\Bigbrk{
\eta^{NR}\eta^{PU}k_1^T
+
\eta^{NT}
\bigbrk{\eta^{RU}k_2^P-\eta^{PU}k_1^R}
}\,.
\end{equation}
Expanding the twisted polarization tensors and performing the index
contractions gives
\begin{equation}
\bigvev{
V_{2,\mathrm{T},k_1}^{(-1,-1)}(z_1,\bar z_1)\,
V_{2,\mathrm{T},k_2}^{(-1,-1)}(z_2,\bar z_2)\,
V_{2,\mathrm{T},k_3}^{(0,0)}(z_3,\bar z_3)
}
\\
=
2\times3^3\times7\times13
\,\gc^3\,\delta\bigbrk{{\textstyle\sum_jk_j}}\,.
\label{eq:V2V2V2twisted}
\end{equation}
The result for untwisted vertices again reduces to the negative of
\eqref{eq:V2V2V2twisted} for $J_i\ll\sqrt{1/\ap}$, plus $\order{J_i^2\ap}$ terms.

\paragraph{\texorpdfstring{$\bigvev{V_1V_3V_3}$}{V1V3V3}.}

Using the VEVs \eqref{eq:corr133t1,eq:corr133t2,eq:corr133t4},
we can compute the correlator
\begin{multline}
\bigvev{
V_{1,k_1}^{(-1,-1)}(z_1,\bar z_1)\,
V_{3,k_2}^{(-1/2,-1/2)}(z_2,\bar z_2)\,
V_{3,k_3}^{(-1/2,-1/2)}(z_3,\bar z_3)
}
\\
=
-\gc^3\,\delta\bigbrk{{\textstyle\sum_jk_j}}
\,\varepsilon^{k_1}_{MN,\tilde M\tilde N}
\,t^{k_2}_{\tilde P\tilde A,PA}
\,t^{k_3}_{\tilde Q\tilde B,QB}
\,X\indup{L}^{MNPQ,AB}X\indup{R}^{\tilde M\tilde N\tilde P\tilde Q,\tilde A\tilde B}\,,
\end{multline}
where
\begin{multline}
X\indup{L,R}^{MNPQ,AB}
=
\frac{\sqrt{\ap}}{2}
\Big[
	\bigbrk{-\eta^{NP}k_1^Q+\eta^{NQ}k_1^P-\eta^{PQ}k_2^N+\sapt k_1^Qk_1^Pk_2^N}
	\Gamma^MC
	\\+
	\bigbrk{\eta^{NQ}+\sapt k_1^Qk_2^N}
	\eta^{MP}\slk_2C
	+
	\bigbrk{-\eta^{NP}+\sapt k_1^Pk_2^N}
	\eta^{MQ}\slk_3C
	-
	\sapt
	\sfrac{1}{2}
	k_2^N
	\eta^{PQ}\slk_2
	\Gamma^M
	\slk_3C
\Big]^{AB}
\end{multline}
are the left/right-moving field contractions, which have been
simplified using \eqref{eq:tgamma0} as well as the symmetry of
$\varepsilon_{k_1}$ and the fact that all contractions of $k_j$ with
$\varepsilon_{k_j}$, $t_{k_j}$ vanish.
Performing the matrix algebra and all the contractions, and using the
twisted polarization tensors $t\indup{T}$, $\varepsilon\indup{T}$, the
correlator evaluates to
\begin{equation}
\bigvev{
V_{1,\mathrm{T},k_1}^{(-1,-1)}(z_1,\bar z_1)\,
V_{3,\mathrm{T},k_2}^{(-1/2,-1/2)}(z_2,\bar z_2)\,
V_{3,\mathrm{T},k_3}^{(-1/2,-1/2)}(z_3,\bar z_3)
}
\\
=
2^8\times5^2
\,\gc^3\,\delta\bigbrk{{\textstyle\sum_jk_j}}\,.
\end{equation}
For untwisted vertices, the correlator is again a lengthy rational expression
in $J_1$, $J_2$, $J_3$.
For $J_i\ll\sqrt{1/\ap}$, it reduces to
\begin{equation}
\bigvev{
V_{1,k_1}^{(-1,-1)}(z_1,\bar z_1)\,
V_{3,k_2}^{(-1/2,-1/2)}(z_2,\bar z_2)\,
V_{3,k_3}^{(-1/2,-1/2)}(z_3,\bar z_3)
}
\\
=
-2^5\times5\times59
\,\gc^3\,\delta\bigbrk{{\textstyle\sum_jk_j}}
+\order{J_i^2\ap}\,.
\end{equation}
%

\paragraph{\texorpdfstring{$\bigvev{V_2V_3V_3}$}{V2V3V3}.}

For the correlator $\bigvev{
V_{2,k_1}^{(-1,-1)}\,
V_{3,k_2}^{(-1/2,-1/2)}\,
V_{3,k_3}^{(-1/2,-1/2)}
}$,
the required fermionic correlators are
\eqref{eq:corr233f1,eq:corr233f2,eq:corr233f4}.
Assembling all terms, and taking
into account the asymmetry of the polarization tensor $\alpha$ as well
as $k\cdot\alpha_{k}=k\cdot t_{k}=0$, the
wanted correlator becomes
\begin{multline}
\bigvev{
V_{2,k_1}^{(-1,-1)}(z_1,\bar z_1)\,
V_{3,k_2}^{(-1/2,-1/2)}(z_2,\bar z_2)\,
V_{3,k_3}^{(-1/2,-1/2)}(z_3,\bar z_3)
}
\\
=
-\gc^3\,\delta\bigbrk{{\textstyle\sum_jk_j}}
\,\alpha^{k_1}_{MNP,\tilde M\tilde N\tilde P}\,
\,t^{k_2}_{\tilde Q\tilde A,QA}\,
\,t^{k_3}_{\tilde R\tilde B,RB}
\,X\indup{L}^{MNPQR,AB}X\indup{R}^{\tilde M\tilde N\tilde P\tilde Q\tilde R,\tilde A\tilde B}\,,
\end{multline}
where
\begin{multline}
X\indup{L,R}^{MNPQR,AB}
=
\sqrt{2}\biggsbrk{
	-\frac{1}{4}\bigbrk{\eta^{QR}-\sapt k_1^Qk_1^R}\Gamma^{MNP}C
	+\frac{3}{4}\apt k_1^Q\eta^{MR}\Gamma^{NP}\slk_3C
	\\
	+\frac{3}{4}\apt k_1^R\eta^{MQ}\slk_2\Gamma^{NP}C
	+\frac{1}{8}\apt\slk_2\Bigbrk{
		-\eta^{QR}\Gamma^{MNP}
		+12\eta^{MR}\eta^{NQ}\Gamma^{P}
		}\slk_3C
}^{AB}
\end{multline}
are the left/right-moving field contractions.
Expanding the polarization tensors, the full
correlator can be computed in the same way as for $\bigvev{V_1V_3V_3}$
above. With twisted polarization tensors, the result is
\begin{equation}
\bigvev{
V_{2,\mathrm{T},k_1}^{(-1,-1)}(z_1,\bar z_1)\,
V_{3,\mathrm{T},k_2}^{(-1/2,-1/2)}(z_2,\bar z_2)\,
V_{3,\mathrm{T},k_3}^{(-1/2,-1/2)}(z_3,\bar z_3)
}
\\
=
-2^8\times3\times7
\,\gc^3\,\delta\bigbrk{{\textstyle\sum_jk_j}}\,.
\end{equation}
For untwisted vertices the result is a lengthy expression again; for
$J_i\ll\sqrt{1/\ap}$, it becomes
\begin{multline}
\bigvev{
V_{2,k_1}^{(-1,-1)}(z_1,\bar z_1)\,
V_{3,k_2}^{(-1/2,-1/2)}(z_2,\bar z_2)\,
V_{3,k_3}^{(-1/2,-1/2)}(z_3,\bar z_3)
}
\\
=
-2^5\times3\times5^2\times7
\,\gc^3\,\delta\bigbrk{{\textstyle\sum_jk_j}}
+\order{J_i^2\ap}\,.
\end{multline}
%

\paragraph{\texorpdfstring{$\bigvev{VVV}$}{VVV}.}

We now can assemble the correlator of three massive scalars. If we
compute the correlators with two or three twisted operators we obtain
that all of them vanish, which is accounted for by a supersymmetric Ward identity.
For three primaries, the correlator becomes
\eqref{eq:vvv}
\begin{multline}
\vev{V_{k_1}(z_1,\bar z_1)V_{k_2}(z_2,\bar z_2)V_{k_3}(z_3,\bar z_3)}
\\
=
\gc^3\,\delta\bigbrk{{\textstyle\sum_jk_j}}
\frac{\Sigma^4}{\Delta_1^4\Delta_2^4\Delta_3^4}
\Bigbrk{
\half\Sigma_2^4+\sfrac{9}{2}\Sigma_4^2+(2\Sigma_2^3-3\Sigma^2\Sigma_2^2+6\Sigma^2\Sigma_4+3\Sigma_2\Sigma_4)\ap\Sigma_4
\\
+\half(3\Sigma^4+7\Sigma_2^2-8\Sigma^2\Sigma_2+6\Sigma_4) (\ap\Sigma_4)^2
-(\Sigma^2-3\Sigma_2)(\ap\Sigma_4)^3+(\ap\Sigma_4)^4
}\,,
\end{multline}
where
\begin{equation}
\Sigma_2=\alpha_1\alpha_2+\alpha_1\alpha_3+\alpha_2\alpha_3\,,
\qquad
\Sigma_4=\alpha_1\alpha_2\alpha_3\Sigma\,,
\end{equation}
and $\Sigma$, $\alpha_j$ are defined in \eqref{eq:alphasigma}.
For $J_i\ll\sqrt{1/\ap}$, the correlator reduces to
\eqref{result}
\begin{equation}
\bigvev{
V_{k_1}(z_1,\bar z_1)
V_{k_2}(z_2,\bar z_2)
V_{k_3}(z_3,\bar z_3)
}
=
\frac{3^8}{2^9}
\,\gc^3\,\delta\bigbrk{{\textstyle\sum_jk_j}}
+\order{J_i^2\ap}\,.
\label{eq:VVVexp}
\end{equation}
%

\section{Mixed Correlators}
\label{sec:MCorrelators}

In the previous section, we checked that all correlators with two or three twisted operators vanish and we also computed the correlators with three untwisted vertices, which correspond to three-point functions of primary operators.

In this section we state the results of correlators with two untwisted vertex operators, which we expect to represent correlation functions of both primary and descendant operators. Starting with the three vertices at the massless level, we have
\begin{equation}
\bigvev{
W_{\mathrm{T},k_1}(z_1,\bar z_1)
W_{k_2}(z_2,\bar z_2)
W_{k_3}(z_3,\bar z_3)
}
=
0\,.
\end{equation}
For one massive vertex and two chiral vertices, we also have cancellations for both correlators
\begin{align}
\bigvev{
W_{k_1}(z_1,\bar z_1)
W_{k_2}(z_2,\bar z_2)
V_{\mathrm{T},{k_3}}(z_3,\bar z_3)
}
&=
0\,,
\\
\bigvev{
W_{\mathrm{T},k_1}(z_1,\bar z_1)
W_{k_2}(z_2,\bar z_2)
V_{k_3}(z_3,\bar z_3)
}
&=
0\,.
\end{align}
In the case of two massive operators and a massless vertex we get
\begin{align}
\bigvev{
V_{\mathrm{T},k_1}(z_1,\bar z_1)
V_{k_2}(z_2,\bar z_2)
W_{k_3}(z_3,\bar z_3)
}
&=
0\,,
\\
\bigvev{
V_{k_1}(z_1,\bar z_1)
V_{k_2}(z_2,\bar z_2)
W_{\mathrm{T},k_3}(z_3,\bar z_3)
}
&=
-\gc^3\,\delta\bigbrk{{\textstyle\sum_jk_j}}
\frac{\al_3^4\Sigma^{4}}{2\Delta_1^{4}\Delta_2^{4}}\,.
\end{align}
Finally, for the case of three massive vertices we have
\begin{align}
\bigvev{
V_{\mathrm{T},k_1}(z_1,\bar z_1)
V_{k_2}(z_2,\bar z_2)
V_{k_3}(z_3,\bar z_3)
}
&=
\gc^3\,\delta\bigbrk{{\textstyle\sum_jk_j}}\frac{\al_1^4\Sigma^{4}}{2\Delta_2^{4}\Delta_3^{4}}\,.
\end{align}
In \secref{sec:results} we explained that there cannot be a supercharge
that annihilates two untwisted vertex operators, so it might at first
seem surprising that four of these correlators actually vanish. Notice
though that all of those include a massless vertex operator $W$, which
corresponds to a chiral operator. Chiral primaries are annihilated not
only by the superconformal charges that in our setup correspond to
$Q^{\pm}$, but also by half of the supergenerators $Q_{\al a}$ and
$\tilde Q_{\dot \alpha}^a$. Massless fields are then annihilated by an
extra eight supercharges, which explains why correlators with at least one
twisted vertex and at least one massless operator are indeed vanishing.

\providecommand{\nbbststyle}{\raggedright\footnotesize\parskip0pt}
\bibliographystyle{nb}
\bibliography{references}

\begin{thebibliography}{10}
\providecommand{\href}[2]{#2}
\providecommand{\arxivref}[2]{\href{http://arxiv.org/abs/#1}{#2}}
\providecommand{\doiref}[2]{\href{http://dx.doi.org/#1}{#2}}
\providecommand{\nbbstauthor}[1]{#1}
\providecommand{\nbbstjournal}[1]{\textsf{#1}}
\providecommand{\nbbsttitle}[1]{\textit{``#1''}}
\providecommand{\nbbsturl}[1]{\texttt{#1}}
\providecommand{\nbbsteprint}[1]{\texttt{#1}}
\providecommand{\nbbststyle}{\raggedright\small\parskip0pt}
\nbbststyle

\bibitem{Beisert:2010jr}
\nbbstauthor{N.~Beisert et~al.},
\nbbsttitle{Review of AdS/CFT Integrability: An Overview},
\nbbstjournal{\doiref{10.1007/s11005-011-0529-2}{Lett.~Math.~Phys.~99,~3~(2012)}},
\nbbsteprint{\arxivref{1012.3982}{arxiv:1012.3982}}.

\bibitem{Gromov:2009zb}
\nbbstauthor{N.~Gromov, V.~Kazakov and P.~Vieira},
\nbbsttitle{Exact Spectrum of Planar {$\mathcal{N}=\mathord{}$4} Supersymmetric
  Yang--Mills Theory: Konishi Dimension at Any Coupling},
\nbbstjournal{\doiref{10.1103/PhysRevLett.104.211601}{Phys.~Rev.~Lett.~104,~211601~(2010)}},
\nbbsteprint{\arxivref{0906.4240}{arxiv:0906.4240}}.

\bibitem{Frolov:2010wt}
\nbbstauthor{S.~Frolov},
\nbbsttitle{Konishi operator at intermediate coupling},
\nbbstjournal{\doiref{10.1088/1751-8113/44/6/065401}{J.~Phys.~A44,~065401~(2011)}},
\nbbsteprint{\arxivref{1006.5032}{arxiv:1006.5032}}.

\bibitem{Gubser:1998bc}
\nbbstauthor{S.~S.~Gubser, I.~R.~Klebanov and A.~M.~Polyakov},
\nbbsttitle{Gauge theory correlators from noncritical string theory},
\nbbstjournal{\doiref{10.1016/S0370-2693(98)00377-3}{Phys.~Lett.~B428,~105~(1998)}},
\nbbsteprint{\arxivref{hep-th/9802109}{hep-th/9802109}}.

\bibitem{Roiban:2009aa}
\nbbstauthor{R.~Roiban and A.~A.~Tseytlin},
\nbbsttitle{Quantum strings in $AdS_5\times S^5$: Strong-coupling corrections
  to dimension of Konishi operator},
\nbbstjournal{\doiref{10.1088/1126-6708/2009/11/013}{JHEP~0911,~013~(2009)}},
\nbbsteprint{\arxivref{0906.4294}{arxiv:0906.4294}}.

\bibitem{Gromov:2011de}
\nbbstauthor{N.~Gromov, D.~Serban, I.~Shenderovich and D.~Volin},
\nbbsttitle{Quantum folded string and integrability: From finite size effects
  to Konishi dimension},
\nbbstjournal{\doiref{10.1007/JHEP08(2011)046}{JHEP~1108,~046~(2011)}},
\nbbsteprint{\arxivref{1102.1040}{arxiv:1102.1040}}.

\bibitem{Beccaria:2011uz}
\nbbstauthor{M.~Beccaria and G.~Macorini},
\nbbsttitle{Quantum Folded String in S$^5$ and the Konishi Multiplet at Strong
  Coupling},
\nbbstjournal{\doiref{10.1007/JHEP10(2011)040}{JHEP~1110,~040~(2011)}},
\nbbsteprint{\arxivref{1108.3480}{arxiv:1108.3480}}.

\bibitem{Roiban:2011fe}
\nbbstauthor{R.~Roiban and A.~A.~Tseytlin},
\nbbsttitle{Semiclassical string computation of strong-coupling corrections to
  dimensions of operators in Konishi multiplet},
\nbbstjournal{\doiref{10.1016/j.nuclphysb.2011.02.016}{Nucl.~Phys.~B848,~251~(2011)}},
\nbbsteprint{\arxivref{1102.1209}{arxiv:1102.1209}}.

\bibitem{Beccaria:2012xm}
\nbbstauthor{M.~Beccaria, S.~Giombi, G.~Macorini, R.~Roiban and
  A.~A.~Tseytlin},
\nbbsttitle{`Short' Spinning Strings and Structure of Quantum
  {AdS$_5\times$S$^5$} Spectrum},
\nbbstjournal{\doiref{10.1103/PhysRevD.86.066006}{Phys.~Rev.~D86,~066006~(2012)}},
\nbbsteprint{\arxivref{1203.5710}{arxiv:1203.5710}}.

\bibitem{Beccaria:2012kp}
\nbbstauthor{M.~Beccaria and A.~A.~Tseytlin},
\nbbsttitle{More About `Short' Spinning Quantum Strings},
\nbbstjournal{\doiref{10.1007/JHEP07(2012)089}{JHEP~1207,~089~(2012)}},
\nbbsteprint{\arxivref{1205.3656}{arxiv:1205.3656}}.

\bibitem{Basso:2010in}
\nbbstauthor{B.~Basso},
\nbbsttitle{Exciting the GKP String at Any Coupling},
\nbbstjournal{\doiref{10.1016/j.nuclphysb.2011.12.010}{Nucl.~Phys.~B857,~254~(2012)}},
\nbbsteprint{\arxivref{1010.5237}{arxiv:1010.5237}}.

\bibitem{Gromov:2011bz}
\nbbstauthor{N.~Gromov and S.~Valatka},
\nbbsttitle{Deeper Look into Short Strings},
\nbbstjournal{\doiref{10.1007/JHEP03(2012)058}{JHEP~1203,~058~(2012)}},
\nbbsteprint{\arxivref{1109.6305}{arxiv:1109.6305}}.

\bibitem{Frolov:2013lva}
\nbbstauthor{S.~Frolov, M.~Heinze, G.~Jorjadze and J.~Plefka},
\nbbsttitle{Static Gauge and Energy Spectrum of Single-Mode Strings in
  {AdS$_5\times$S$^5$}},
\nbbstjournal{\doiref{10.1088/1751-8113/47/8/085401}{J.~Phys.~A47,~085401~(2014)}},
\nbbsteprint{\arxivref{1310.5052}{arxiv:1310.5052}}.

\bibitem{Escobedo:2010xs}
\nbbstauthor{J.~Escobedo, N.~Gromov, A.~Sever and P.~Vieira},
\nbbsttitle{Tailoring Three-Point Functions and Integrability},
\nbbstjournal{\doiref{10.1007/JHEP09(2011)028}{JHEP~1109,~028~(2011)}},
\nbbsteprint{\arxivref{1012.2475}{arxiv:1012.2475}}.

\bibitem{Escobedo:2011xw}
\nbbstauthor{J.~Escobedo, N.~Gromov, A.~Sever and P.~Vieira},
\nbbsttitle{Tailoring Three-Point Functions and Integrability II. Weak/strong
  coupling match},
\nbbstjournal{\doiref{10.1007/JHEP09(2011)029}{JHEP~1109,~029~(2011)}},
\nbbsteprint{\arxivref{1104.5501}{arxiv:1104.5501}}.

\bibitem{Gromov:2011jh}
\nbbstauthor{N.~Gromov, A.~Sever and P.~Vieira},
\nbbsttitle{Tailoring Three-Point Functions and Integrability III. Classical
  Tunneling},
\nbbstjournal{\doiref{10.1007/JHEP07(2012)044}{JHEP~1207,~044~(2012)}},
\nbbsteprint{\arxivref{1111.2349}{arxiv:1111.2349}}.

\bibitem{Kostov:2012yq}
\nbbstauthor{I.~Kostov},
\nbbsttitle{Three-point function of semiclassical states at weak coupling},
\nbbsteprint{\arxivref{1205.4412}{arxiv:1205.4412}}.

\bibitem{Foda:2013nua}
\nbbstauthor{O.~Foda, Y.~Jiang, I.~Kostov and D.~Serban},
\nbbsttitle{A Tree-Level 3-Point Function in the $su(3)$-Sector of Planar
  {$\mathcal{N}=\mathord{}$4} SYM},
\nbbstjournal{\doiref{10.1007/JHEP10(2013)138}{JHEP~1310,~138~(2013)}},
\nbbsteprint{\arxivref{1302.3539}{arxiv:1302.3539}}.

\bibitem{Zarembo:2010rr}
\nbbstauthor{K.~Zarembo},
\nbbsttitle{Holographic three-point functions of semiclassical states},
\nbbstjournal{\doiref{10.1007/JHEP09(2010)030}{JHEP~1009,~030~(2010)}},
\nbbsteprint{\arxivref{1008.1059}{arxiv:1008.1059}}.

\bibitem{Costa:2010rz}
\nbbstauthor{M.~S.~Costa, R.~Monteiro, J.~E.~Santos and D.~Zoakos},
\nbbsttitle{On three-point correlation functions in the gauge/gravity duality},
\nbbstjournal{\doiref{10.1007/JHEP11(2010)141}{JHEP~1011,~141~(2010)}},
\nbbsteprint{\arxivref{1008.1070}{arxiv:1008.1070}}.

\bibitem{Minahan:2012fh}
\nbbstauthor{J.~A.~Minahan},
\nbbsttitle{Holographic three-point functions for short operators},
\nbbstjournal{\doiref{10.1007/JHEP07(2012)187}{JHEP~1207,~187~(2012)}},
\nbbsteprint{\arxivref{1206.3129}{arxiv:1206.3129}}.

\bibitem{Polchinski:1999ry}
\nbbstauthor{J.~Polchinski},
\nbbsttitle{S matrices from AdS space-time},
\nbbsteprint{\arxivref{hep-th/9901076}{hep-th/9901076}}.

\bibitem{Susskind:1998vk}
\nbbstauthor{L.~Susskind},
\nbbsttitle{Holography in the flat space limit},
\nbbsteprint{\arxivref{hep-th/9901079}{hep-th/9901079}}.

\bibitem{Balasubramanian:1999ri}
\nbbstauthor{V.~Balasubramanian, S.~B.~Giddings and A.~E.~Lawrence},
\nbbsttitle{What do CFTs tell us about Anti-de Sitter space-times?},
\nbbstjournal{\doiref{10.1088/1126-6708/1999/03/001}{JHEP~9903,~001~(1999)}},
\nbbsteprint{\arxivref{hep-th/9902052}{hep-th/9902052}}.

\bibitem{Bianchi:2003wx}
\nbbstauthor{M.~Bianchi, J.~F.~Morales and H.~Samtleben},
\nbbsttitle{On stringy {$AdS_5\times S^5$} and higher spin holography},
\nbbstjournal{\doiref{10.1088/1126-6708/2003/07/062}{JHEP~0307,~062~(2003)}},
\nbbsteprint{\arxivref{hep-th/0305052}{hep-th/0305052}}.

\bibitem{Beisert:2003te}
\nbbstauthor{N.~Beisert, M.~Bianchi, J.~F.~Morales and H.~Samtleben},
\nbbsttitle{On the Spectrum of AdS/CFT beyond Supergravity},
\nbbstjournal{\doiref{10.1088/1126-6708/2004/02/001}{JHEP~0402,~001~(2004)}},
\nbbsteprint{\arxivref{hep-th/0310292}{hep-th/0310292}}.

\bibitem{Boels:2012if}
\nbbstauthor{R.~H.~Boels},
\nbbsttitle{Three particle superstring amplitudes with massive legs},
\nbbstjournal{\doiref{10.1007/JHEP06(2012)026}{JHEP~1206,~026~(2012)}},
\nbbsteprint{\arxivref{1201.2655}{arxiv:1201.2655}}.

\bibitem{Lee:1998bxa}
\nbbstauthor{S.~Lee, S.~Minwalla, M.~Rangamani and N.~Seiberg},
\nbbsttitle{Three point functions of chiral operators in {$D=4$},
  {$\mathcal{N}=\mathord{}$4} SYM at large N},
\nbbstjournal{Adv.~Theor.~Math.~Phys.~2,~697~(1998)},
\nbbsteprint{\arxivref{hep-th/9806074}{hep-th/9806074}}.

\bibitem{Klose:2011rm}
\nbbstauthor{T.~Klose and T.~McLoughlin},
\nbbsttitle{A Light-Cone Approach to Three-Point Functions in $AdS_5\times
  S^5$},
\nbbstjournal{\doiref{10.1007/JHEP04(2012)080}{JHEP~1204,~080~(2012)}},
\nbbsteprint{\arxivref{1106.0495}{arxiv:1106.0495}}.

\bibitem{Witten:1998qj}
\nbbstauthor{E.~Witten},
\nbbsttitle{Anti-de~Sitter space and holography},
\nbbstjournal{Adv.~Theor.~Math.~Phys.~2,~253~(1998)},
\nbbsteprint{\arxivref{hep-th/9802150}{hep-th/9802150}}.

\bibitem{Freedman:1998tz}
\nbbstauthor{D.~Z.~Freedman, S.~D.~Mathur, A.~Matusis and L.~Rastelli},
\nbbsttitle{Correlation functions in the CFT(d)/AdS(d+1) correspondence},
\nbbstjournal{\doiref{10.1016/S0550-3213(99)00053-X}{Nucl.~Phys.~B546,~96~(1999)}},
\nbbsteprint{\arxivref{hep-th/9804058}{hep-th/9804058}}.

\bibitem{Buchbinder:2011jr}
\nbbstauthor{E.~I.~Buchbinder and A.~A.~Tseytlin},
\nbbsttitle{Semiclassical correlators of three states with large $S^5$ charges
  in string theory in $AdS_5\times S^5$},
\nbbstjournal{\doiref{10.1103/PhysRevD.85.026001}{Phys.~Rev.~D85,~026001~(2012)}},
\nbbsteprint{\arxivref{1110.5621}{arxiv:1110.5621}}.

\bibitem{Friedan:1985ge}
\nbbstauthor{D.~Friedan, E.~J.~Martinec and S.~H.~Shenker},
\nbbsttitle{Conformal Invariance, Supersymmetry and String Theory},
\nbbstjournal{Nucl.~Phys.~B271,~93~(1986)}.

\bibitem{Koh:1987hm}
\nbbstauthor{I.~G.~Koh, W.~Troost and A.~Van~Proeyen},
\nbbsttitle{Covariant Higher Spin Vertex Operators in the Ramond Sector},
\nbbstjournal{\doiref{10.1016/0550-3213(87)90642-0}{Nucl.~Phys.~B292,~201~(1987)}}.

\bibitem{Beisert:2002tn}
\nbbstauthor{N.~Beisert},
\nbbsttitle{BMN Operators and Superconformal Symmetry},
\nbbstjournal{\doiref{10.1016/S0550-3213(03)00229-3}{Nucl.~Phys.~B659,~79~(2003)}},
\nbbsteprint{\arxivref{hep-th/0211032}{hep-th/0211032}}.

\bibitem{Minahan:2002ve}
\nbbstauthor{J.~A.~Minahan and K.~Zarembo},
\nbbsttitle{The Bethe-ansatz for {$\mathcal{N}=\mathord{}$4} super
  Yang--Mills},
\nbbstjournal{\doiref{10.1088/1126-6708/2003/03/013}{JHEP~0303,~013~(2003)}},
\nbbsteprint{\arxivref{hep-th/0212208}{hep-th/0212208}}.

\bibitem{D'Hoker:1999ea}
\nbbstauthor{E.~D'Hoker, D.~Z.~Freedman, S.~D.~Mathur, A.~Matusis and
  L.~Rastelli},
\nbbsttitle{Extremal correlators in the {AdS/CFT} correspondence},
\nbbsteprint{\arxivref{hep-th/9908160}{hep-th/9908160}},
in: \nbbsttitle{The Many Faces of the Superworld},
ed.: M.~Shifman,
World Scientific (1999),
Singapore.

\bibitem{Janik:2010gc}
\nbbstauthor{R.~A.~Janik, P.~Surowka and A.~Wereszczynski},
\nbbsttitle{On correlation functions of operators dual to classical spinning
  string states},
\nbbstjournal{\doiref{10.1007/JHEP05(2010)030}{JHEP~1005,~030~(2010)}},
\nbbsteprint{\arxivref{1002.4613}{arxiv:1002.4613}}.

\bibitem{D'Hoker:1999pj}
\nbbstauthor{E.~D'Hoker, D.~Z.~Freedman, S.~D.~Mathur, A.~Matusis and
  L.~Rastelli},
\nbbsttitle{Graviton Exchange and Complete Four Point Functions in the AdS/CFT
  Correspondence},
\nbbstjournal{\doiref{10.1016/S0550-3213(99)00525-8}{Nucl.~Phys.~B562,~353~(1999)}},
\nbbsteprint{\arxivref{hep-th/9903196}{hep-th/9903196}}.

\bibitem{D'Hoker:1999ni}
\nbbstauthor{E.~D'Hoker, D.~Z.~Freedman and L.~Rastelli},
\nbbsttitle{AdS/CFT Four Point Functions: How to Succeed at z Integrals Without
  Really Trying},
\nbbstjournal{\doiref{10.1016/S0550-3213(99)00526-X}{Nucl.~Phys.~B562,~395~(1999)}},
\nbbsteprint{\arxivref{hep-th/9905049}{hep-th/9905049}}.

\bibitem{Penedones:2010ue}
\nbbstauthor{J.~Penedones},
\nbbsttitle{Writing CFT Correlation Functions as AdS Scattering Amplitudes},
\nbbstjournal{\doiref{10.1007/JHEP03(2011)025}{JHEP~1103,~025~(2011)}},
\nbbsteprint{\arxivref{1011.1485}{arxiv:1011.1485}}.

\bibitem{Fitzpatrick:2011ia}
\nbbstauthor{A.~L.~Fitzpatrick, J.~Kaplan, J.~Penedones, S.~Raju and
  B.~C.~van~Rees},
\nbbsttitle{A~Natural Language for AdS/CFT Correlators},
\nbbstjournal{\doiref{10.1007/JHEP11(2011)095}{JHEP~1111,~095~(2011)}},
\nbbsteprint{\arxivref{1107.1499}{arxiv:1107.1499}}.

\bibitem{Costa:2012cb}
\nbbstauthor{M.~S.~Costa, V.~Goncalves and J.~Penedones},
\nbbsttitle{Conformal Regge theory},
\nbbstjournal{\doiref{10.1007/JHEP12(2012)091}{JHEP~1212,~091~(2012)}},
\nbbsteprint{\arxivref{1209.4355}{arxiv:1209.4355}}.

\bibitem{Arutyunov:2007tc}
\nbbstauthor{G.~Arutyunov and S.~Frolov},
\nbbsttitle{On String S-matrix, Bound States and TBA},
\nbbstjournal{\doiref{10.1088/1126-6708/2007/12/024}{JHEP~0712,~024~(2007)}},
\nbbsteprint{\arxivref{0710.1568}{arxiv:0710.1568}}.

\bibitem{Arutyunov:2009ur}
\nbbstauthor{G.~Arutyunov and S.~Frolov},
\nbbsttitle{Thermodynamic Bethe Ansatz for the $AdS_5\times S^5$ Mirror Model},
\nbbstjournal{\doiref{10.1088/1126-6708/2009/05/068}{JHEP~0905,~068~(2009)}},
\nbbsteprint{\arxivref{0903.0141}{arxiv:0903.0141}}.

\bibitem{Gromov:2009tv}
\nbbstauthor{N.~Gromov, V.~Kazakov and P.~Vieira},
\nbbsttitle{Exact Spectrum of Anomalous Dimensions of Planar
  {$\mathcal{N}=\mathord{}$4} Supersymmetric Yang--Mills Theory},
\nbbstjournal{\doiref{10.1103/PhysRevLett.103.131601}{Phys.~Rev.~Lett.~103,~131601~(2009)}},
\nbbsteprint{\arxivref{0901.3753}{arxiv:0901.3753}}.

\bibitem{Bombardelli:2009ns}
\nbbstauthor{D.~Bombardelli, D.~Fioravanti and R.~Tateo},
\nbbsttitle{Thermodynamic Bethe Ansatz for planar AdS/CFT: a proposal},
\nbbstjournal{\doiref{10.1088/1751-8113/42/37/375401}{J.~Phys.~A42,~375401~(2009)}},
\nbbsteprint{\arxivref{0902.3930}{arxiv:0902.3930}}.

\bibitem{Gromov:2011cx}
\nbbstauthor{N.~Gromov, V.~Kazakov, S.~Leurent and D.~Volin},
\nbbsttitle{Solving the AdS/CFT Y-system},
\nbbstjournal{\doiref{10.1007/JHEP07(2012)023}{JHEP~1207,~023~(2012)}},
\nbbsteprint{\arxivref{1110.0562}{arxiv:1110.0562}}.

\bibitem{Gromov:2013pga}
\nbbstauthor{N.~Gromov, V.~Kazakov, S.~Leurent and D.~Volin},
\nbbsttitle{Quantum Spectral Curve for $AdS_5/CFT_4$},
\nbbstjournal{\doiref{10.1103/PhysRevLett.112.011602}{Phys.~Rev.~Lett.~112,~011602~(2014)}},
\nbbsteprint{\arxivref{1305.1939}{arxiv:1305.1939}}.

\bibitem{Polchinski:1998stringboth}
\nbbstauthor{J.~A.~Polchinski},
\nbbsttitle{String Theory},
Cambridge University Press (1998),
Cambridge, UK.

\bibitem{Kostelecky:1986xg}
\nbbstauthor{V.~A.~Kostelecky, O.~Lechtenfeld, W.~Lerche, S.~Samuel and
  S.~Watamura},
\nbbsttitle{Conformal Techniques, Bosonization and Tree Level String
  Amplitudes},
\nbbstjournal{\doiref{10.1016/0550-3213(87)90213-6}{Nucl.~Phys.~B288,~173~(1987)}}.

\bibitem{Feng:2012bb}
\nbbstauthor{W.-Z.~Feng, D.~L{\"u}st and O.~Schlotterer},
\nbbsttitle{Massive Supermultiplets in Four-Dimensional Superstring Theory},
\nbbstjournal{\doiref{10.1016/j.nuclphysb.2012.03.010}{Nucl.~Phys.~B861,~175~(2012)}},
\nbbsteprint{\arxivref{1202.4466}{arxiv:1202.4466}}.

\end{thebibliography}

\end{document}